\documentclass[aps, prx, twocolumn, showpacs, floatfix,10pt,superscriptaddress]{revtex4-2}
\usepackage{amsmath}
\usepackage{amsfonts}
\usepackage{amsbsy}
\usepackage{amssymb}
\usepackage{graphicx}
\usepackage{textcomp}
\usepackage[caption=false,singlelinecheck=false]{subfig}
\usepackage{xcolor}
\usepackage{mathrsfs}
\usepackage{mathtools}
\usepackage{bm}
\usepackage{gensymb}
\usepackage{braket,wasysym}
\usepackage{float}
\usepackage{tikz}
\usepackage[colorlinks,linkcolor=blue,citecolor=red,filecolor=magenta,urlcolor=red,breaklinks]{hyperref}
\usepackage[bottom]{footmisc}
\usepackage{verbatim}

\hypersetup{colorlinks=true, urlcolor=blue, citecolor=red, pdfborder={0 0 0}}
\usepackage{breakurl}
\usepackage{natbib}

\allowdisplaybreaks

\newcommand{\tcm}[1]{\textcolor{magenta}{#1}}

\normalfont
\def\be{\begin{equation}}
	\def\ee{\end{equation}}
\def\bea{\begin{eqnarray}}
	\def\eea{\end{eqnarray}}

\newcommand{\Z}{\mathbb{Z}}
\newcommand{\R}{\mathbb{R}}
\newcommand{\N}{\mathbb{N}}
\newcommand{\Torus}{\mathbb{T}}
\newcommand{\normZ}[1]{\left\lVert #1\right\rVert_{\Z}}

\begin{document}

\title{Irrationality Measure Controls Long-Wavelength Charge Fluctuations
\\in Quasiperiodic Systems}

\author{Junmo Jeon}
\email{junmojeon@sophia.ac.jp}
\affiliation{Physics Division, Sophia University, Chiyoda-ku, Tokyo 102-8554, Japan}
\author{Shiro Sakai}
\email{shirosakai@sophia.ac.jp}
\affiliation{Physics Division, Sophia University, Chiyoda-ku, Tokyo 102-8554, Japan}

\date{\today}

\begin{abstract}
Quasiperiodic order is characterized by irrational frequencies whose rational approximability known as irrationality measure defines distinct arithmetic classes. We establish that this arithmetic classification has direct physical consequences for long-wavelength charge fluctuations. In translation-covariant quasiperiodic systems, the infrared scaling of charge fluctuation is governed by the interplay between the irrationality exponent of irrational frequency and the large-harmonic decay of the hull charge profile: the latter determines the available charge weight, while the former controls how efficiently that weight is transferred to the infrared. Consequently, all algebraic irrational frequencies share the same arithmetic scaling, whereas exceptionally well-approximable transcendental frequencies can exhibit strongly enhanced infrared fluctuation scaling. We further prove that occupied states separated from the Fermi level by a gap stable throughout the hull contribute only an analytic infrared background, leaving the nontrivial scaling to near-Fermi states. Our results extend to general translation-covariant multi-frequency quasiperiodic systems.
\end{abstract}

\maketitle

\tcm{\textit{Introduction--}}
Quasiperiodic systems realize long-range order without a unit cell, generating structure over an unbounded hierarchy of length scales~\cite{Janot1992,BaakeGrimm2013}. This hierarchy has an intrinsic arithmetic origin: an irrational modulation frequency is accompanied by infinitely many rational approximants, which determine how closely the system can resemble a periodic crystal on progressively larger scales~\cite{Janot1992,BaakeGrimm2013,GoldmanKelton1993,Khinchin1997}. Arithmetic properties of incommensurate frequencies affect critical scaling and long-wavelength fluctuations of quasiperiodic point sets~\cite{PhysRevB.98.134201,BjorklundHartnick2024}, yet their role in electronic properties remains unclear.

Despite the absence of periodicity, charge densities in quasiperiodic systems exhibit suppressed long-wavelength fluctuations characteristic of hyperuniformity~\cite{TorquatoStillinger2003,Torquato2018,PhysRevX.11.021028}, distinguishing them from disordered systems~\cite{Sakai22PRB,SakaiAritaOhtsuki2022,Hori24,XiangJeonSakai2026}. Recent studies have classified this behavior using specific models, such as the Aubry--Andr\'e model~\cite{AubryAndre1980} with the golden ratio frequency, and linked the resulting scaling to the spectral and localization properties near the Fermi level~\cite{SakaiAritaOhtsuki2022,XiangJeonSakai2026}. These studies found regimes with either asymptotically bounded density fluctuations or logarithmically growing fluctuations~\cite{SakaiAritaOhtsuki2022,XiangJeonSakai2026}. However, since such conclusions are based on a limited set of models and irrational frequencies, the general principle governing long-wavelength charge fluctuations in quasiperiodic systems remains unknown.



In this Letter, we show that the intrinsic arithmetic property of the irrational frequency known as irrationality measure plays the role of a fundamental scaling variable governing long-wavelength charge fluctuations in quasiperiodic systems. Near-Fermi physics determines which quasiperiodic components carry appreciable weight, while arithmetic controls how efficiently this weight can be transferred to the infrared. All algebraic irrational frequencies share the same arithmetic scaling property, while some transcendental irrational frequencies with larger irrationality exponents approach commensurability exceptionally closely (i.e., resonances), allowing appreciable weight to reach smaller wavevectors. Occupied states separated from the Fermi level by a stable gap contribute only a smooth analytic background, whereas the near-Fermi states determine the degree of the long-range fluctuations. Here, exceptionally well approximable irrational frequencies can generate fluctuations stronger than logarithmic growth and can even destroy hyperuniformity. This mechanism is general and extends naturally to multi-frequency quasiperiodic systems.

\tcm{\textit{Hyperuniformity framework--}}
Let us briefly review the real- and reciprocal-space diagnostics of hyperuniformity used below~\cite{TorquatoStillinger2003,Torquato2018}. Consider a scalar field $n_j$ on one-dimensional lattice sites $j\in\Z$. For a window of radius $R$ centered at a position $x_c\in\R$, we define the enclosed charge by
\begin{equation}
 N_R(x_c)=\sum_{j\in\Z}n_j\,\Theta(R-|j-x_c|),
 \label{eq:windowcharge}
\end{equation}
where $\Theta$ is the Heaviside step function, and its variance,
\begin{equation}
 \sigma^2(R)=\overline{N_R(x_c)^2}-\overline{N_R(x_c)}^{\,2},
 \label{eq:variance}
\end{equation}
where the overline denotes an average over $x_c$.
In one dimension, hyperuniformity means that $\sigma^2(R)=o(R)$ for large $R$. If $\sigma^2(R)/R$ does not vanish for $R\to\infty$, the distribution is nonhyperuniform. An infrared increasing of $\sigma^2(R)/R$ is referred to as anti-hyperuniform.
The large-$R$ behavior of $ \sigma^2(R)$ of hyperuniform distribution further distinguishes the following classes: $\sigma^2(R)$ approaches a constant for class I, grows as $\ln R$ for class II, and grows algebraically for class III. A convenient diagnostic is the ratio $\sigma^2/\ln R$ that tends to zero for class I, approaches a positive constant for class II, and increases with $R$ for class III~\cite{TorquatoStillinger2003,Torquato2018,SM}.

In reciprocal space, long-wavelength charge fluctuations are characterized by the structure factor $S(k)$. Since quasiperiodic diffraction generally consists of Bragg peaks rather than a continuous spectrum, we use the integrated infrared weight
\begin{equation}
 Z(K)=\int_{|k|<K}S(k)\,dk.
 \label{eq:Zdef}
\end{equation}
For an ordinary power law $S(k)\sim |k|^\alpha$, one has $Z(K)\sim K^{\alpha+1}$: $\alpha>1$, $\alpha=1$, and $0<\alpha<1$ correspond to classes I, II, and III, respectively, while $\alpha=0$ and $\alpha<0$ correspond to non-hyperuniform and anti-hyperuniform behaviors~\cite{PhysRevB.95.054119}. For the pure-point quasiperiodic spectrum, we define the infrared envelope exponent by
\begin{equation}
 \alpha+1:=\liminf_{K\to0^+}\frac{\log Z(K)}{\log K}.
 \label{eq:alphadef}
\end{equation}
At the marginal value $\alpha=0$, the logarithmic exponent in Eq.~\eqref{eq:alphadef} alone does not determine hyperuniformity, and the subleading behavior of $Z(K)/K$ must also be examined. 

Previous studies of the Aubry--Andr\'e~\cite{AubryAndre1980,SakaiAritaOhtsuki2022} and generalized Aubry--Andr\'e models~\cite{GaneshanPixleyDasSarma2015,XiangJeonSakai2026}, using the conventional golden-ratio [$\varphi=(\sqrt{5}-1)/2$] quasiperiodicity, found a strikingly simple relation between the many-body charge-density distribution and single-particle states around the Fermi level: gapped phases and gapless phases with extended states at the Fermi energy give a class-I hyperuniform charge distribution, whereas gapless phases with localized or critical states at the Fermi energy give a class-II distribution. These observations indicate that the long-wavelength charge fluctuations are controlled primarily by low-energy states near the Fermi level rather than by the entire occupied sea. We now show that this picture is both more general and less universal than it first appears. 

Before going into details, let us briefly summarize our main result. We find that long-wavelength charge fluctuations in general translation-covariant quasiperiodic systems are controlled not only by the localization characteristics of single-particle states near the Fermi level, but also by the irrationality exponent of quasiperiodic frequency of the Hamiltonian itself. In this sense, the fundamental nature of irrationality encoded in the quasiperiodic Hamiltonian itself acts as a physical scaling variable [Eq.~\eqref{eq:main} below]. The conventional golden-ratio quasiperiodicity belongs to the usual arithmetic sector and naturally reproduces the familiar class-I/II behavior, whereas exceptionally well approximable irrational frequencies with larger irrationality exponent can generate stronger infrared fluctuations, giving class III and even nonhyperuniform or anti-hyperuniform distributions. We also show rigorously that occupied states separated from the Fermi level by a gap do not affect the nontrivial infrared scaling, as far as the gap remains open under changes of the dynamical hull phase.
This explains why the state at the Fermi level determines the long-wavelength behavior of the charge distribution in the previous studies~\cite{SakaiAritaOhtsuki2022,XiangJeonSakai2026}. Now we formulate these results for a general translation-covariant quasiperiodic Hamiltonian.

\tcm{\textit{Hull harmonics and arithmetic down-conversion--}}
Let $\{|j\rangle:j\in\Z\}$ be the localized site basis, and let $T$ be the one-site translation operator defined by $T|j\rangle=|j+1\rangle$. We consider a translation-covariant family of Hamiltonians $H(\theta)$ defined over a $2\pi$-periodic hull phase $\theta$~\cite{ProdanSchulzBaldes2016,BellissardHerrmannZarrouati2000,BaakeGrimm2013}, satisfying
\begin{equation}
T H(\theta)T^{-1}=H(\theta-2\pi\beta).
\label{eq}
\end{equation}
Here, $\beta\notin\mathbb Q$ denotes the irrational modulation frequency which can be replaced by a rational approximant whose denominator is the system size $L$. Thus, shifting the lattice by one site is equivalent to advancing the hull phase by $2\pi\beta$.

The Aubry--Andr\'e model,
\begin{equation}
 \begin{split}
 H(\theta)=&-t\sum_{j\in\Z}(f_j^\dagger f_{j+1}+{\rm H.c.})+\lambda\sum_{j\in\Z}\cos(2\pi\beta j+\theta)f_j^\dagger f_j 
 \end{split}
 \label{eq:AA}
\end{equation}
is a canonical example.
The operators $f_j^\dagger$ ($f_j$) creates (annihilates) a spinless fermion at site $j$. $t$ is the nearest-neighbor hopping amplitude, and $\lambda$ is the potential strength. For a general Diophantine irrational $\beta$, the Aubry--Andr\'e model exhibits a self-dual localization transition at $\lambda=2t$. The single-particle eigenstates are extended for $\lambda<2t$, critical (neither localized nor extended) at $\lambda=2t$, and exponentially localized for $\lambda>2t$~\cite{AubryAndre1980,Jitomirskaya1999}.

Now let us consider the zero-temperature Fermi projector $P_F(\theta)=\mathbf 1_{(-\infty,E_F]}[H(\theta)]$, where $E_F$ is the Fermi energy, and define the hull charge profile $F(\vartheta):=\langle0|P_F(\vartheta)|0\rangle$. Translation covariance gives the local density $n_j(\theta):=\langle j|P_F(\theta)|j\rangle=F(\theta+2\pi\beta j)$. We write
\begin{equation}
 F(\vartheta)-\overline F=\sum_{m\ne0}c_m e^{im\vartheta},
 \label{eq:hull}
\end{equation}
where $\overline F=(2\pi)^{-1}\int_0^{2\pi}F(\vartheta)\,d\vartheta$ is the hull average and $c_m$ is the $m$-th Fourier coefficient. 
The same representation holds for any translation-covariant local scalar observable.

For a real number $x$, let $\normZ{x}:=\min_{\ell\in\Z}|x-\ell|$ denote its distance to the nearest integer. Modulo reciprocal-lattice vectors, the magnitude of the physical wave number carried by the $m$-th hull harmonic is
\begin{equation}
 k_m=2\pi\normZ{m\beta}.
 \label{eq:qmap}
\end{equation}
Note that the hull profile determines $|c_m|^2$, whereas the distance of that weight from zero momentum is determined entirely by arithmetic of $\beta$. This distance is related to the rational approximability of irrational freqeuncy of quasiperiodic Hamiltonian known as irrationality measure.

We quantify irrationality measure by the standard irrationality exponent~\cite{Bugeaud2004,Roth1955}
\begin{equation}
 \begin{aligned}
 \mu(\beta)=\sup\{\nu:\;&\left|\beta-\frac pq\right|<q^{-\nu}\\
 &\text{for infinitely many }(p,q)\in\Z\times\N\}.
 \end{aligned}
 \label{eq:mu}
\end{equation}
The physical resonance exponent is
\begin{equation}
 \omega(\beta):=\limsup_{m\to\infty}\frac{-\log\normZ{m\beta}}{\log m}.
 \label{eq:omega}
\end{equation}
Since $|\beta-p/q|=\normZ{q\beta}/q$ when $p$ is the nearest integer to $q\beta$, one has $\omega(\beta)=\mu(\beta)-1$. Hence a large irrationality exponent creates a sparse hierarchy of denominators for which translation by $q$ sites produces an exceptionally small phase mismatch. Those are precisely the scales at which high hull harmonics are down-converted to anomalously small physical momenta~\cite{BaakeGrimm2013,BjorklundHartnick2024}.

\tcm{\textit{Arithmetic control of the scaling exponent--}}
Suppose the relevant hull Fourier amplitudes obey the algebraic upper bound,
\begin{equation}
|c_m|\lesssim |m|^{-r},
\label{eq:cmupperbound}
\end{equation}
for $r>1/2$ and retain weight of the same order along a subsequence of resonances realizing the irrationality exponent $\omega(\beta)$, which we refer to as an exponent-realizing sequence,  $|c_{q_j}|\gtrsim  q_j^{-r}$.
This is only a mild nondegeneracy condition since it excludes a systematic cancellation of the strongest arithmetic resonances by zeros of the observable, rather than requiring nonvanishing weight at all large harmonics.
The arithmetic of $\beta$ determines which hull harmonics are transferred to exceptionally small momenta, while $E_F$ controls their Fourier weights. Hence, accidental cancellations at special fillings are generically lifted by an arbitrarily small shift of $E_F$~\cite{SM}. 
Then, Eq.~\eqref{eq:Zdef} reduces, up to an inessential normalization, to a sum over the harmonics $\vert c_m\vert^2$ satisfying $\normZ{m\beta}<K$.

Our central result is the 
relation,
\begin{equation}
 \alpha+1=\frac{2r}{\mu(\beta)-1}
 \label{eq:main}
\end{equation}
for $2\le\mu(\beta)<\infty$.
The physical origin of Eq.~\eqref{eq:main} is as follows. For an exceptionally good rational approximant $p/q$ for a given irrational $\beta$, the $q$-th hull harmonic is folded to a physical momentum
$k_q\sim \lVert q\beta\rVert_{\mathbb Z}\sim q^{-(\mu-1)}$,
while its charge weight scales as
$|c_q|^2\sim q^{-2r}$.
Eliminating $q$ therefore gives a low-momentum weight proportional to
$k_q^{\,2r/(\mu-1)}$, suggesting Eq.~\eqref{eq:main}.
To establish this scaling rigorously, one should also control all other harmonics entering the same infrared window. Two such resonant Fourier indices have a difference $d$ satisfying
$\lVert \beta d\rVert_{\mathbb Z}<2K$; the irrationality exponent then forces
$d\gtrsim K^{-1/(\mu-1+\epsilon)}$ for any $\epsilon>0$.
The resonant indices are therefore sparse in Fourier space of $\theta$. Hence, their algebraically decaying weights cannot collectively generate a stronger infrared singularity. Combining this upper bound with the weight of an exponent-realizing exceptional subsequence yields Eq.~\eqref{eq:main}. The full proof is presented in the Supplemental Material~\cite{SM}.

Equation~\eqref{eq:main} reveals an unusual relation between order and fluctuations. Increasing $r$ removes high-harmonic weight and strengthens hyperuniformity. Increasing $\mu$ has the opposite effect: the system becomes extraordinarily close to a periodic approximant at rare scales, yet this stronger near-periodicity repeatedly pushes finite Fourier weight to smaller momenta. Hence, better rational approximability produces stronger fluctuations at longer distances.

Notably, the exponents $r$ and $\mu$ encode distinct physical ingredients in Eq.~\eqref{eq:main}. The irrationality exponent $\mu$ is a purely arithmetic property of the irrational frequency defining the quasiperiodic Hamiltonian and is independent of the Fermi energy. Larger $\mu$ corresponds to stronger near-commensurability over rare long length scales, which lowers $\alpha$ and enhances long-wavelength fluctuations. In contrast, $r$ reflects the functional regularity of the hull charge profile which depends on the Fermi level. A smoother hull profile gives larger $r$, increases $\alpha$, and suppresses long-wavelength fluctuations.

The arithmetic distinction is particularly sharp in Eq.~\eqref{eq:main}.
From the Thue–Siegel–Roth theorem~\cite{Roth1955}, all algebraic irrational numbers including golden ratio $\varphi$ have $\mu=2$, and hence they have $\alpha=2r-1$. Any irrational numbers with $\mu>2$ are necessarily transcendental and can exhibit parametrically stronger exceptional resonances. The converse is false: transcendental numbers with $\mu=2$ remain in the same arithmetic scaling sector as algebraic irrationals. Thus, the observable distinction is the Diophantine class itself, with algebraicity providing a universal subset.

Several useful corollaries follow directly from the smoothness of the hull charge profile $F(\theta)$. First, if $F$ is $s$ times continuously differentiable with respect to $\theta$, its Fourier coefficients decay at least as $c_m=o(|m|^{-s})$. Thus, an algebraic tail $|c_m|\sim |m|^{-r}$ should satisfy $r\ge s$. Combining this regularity bound with Eq.~\eqref{eq:main}, one finds that the inequality $s>\mu(\beta)-1$, which generally holds for extended gapless phases, is sufficient to guarantee class-I hyperuniformity ($\alpha>1$). Physically, sufficiently smooth variation of the charge density over the hull suppresses high-order harmonics strongly enough that even anomalously strong arithmetic resonances cannot generate class-II or -III long-wavelength fluctuations.
Second, if $F$ has a simple jump and is otherwise sufficiently regular, its Fourier amplitudes have a $1/|m|$ tail, giving $r=1$. For multiple jumps and piecewise regular $F$, the same conclusion applies provided the leading Fourier weight does not cancel along the $\mu$ exponent-realizing resonances. Then, Eq.~\eqref{eq:main} gives $\alpha=2/[\mu(\beta)-1]-1$. Thus, $\mu=2$ yields a class II, while $2<\mu<3$ gives a class III. Such behaviors are widely found in the critical and localized gapless phases due to the strong inhomogeneity of wavefunctions~\cite{SakaiAritaOhtsuki2022,XiangJeonSakai2026}. 

\begin{figure}[h]
    \centering
    \includegraphics[width=0.45\textwidth]{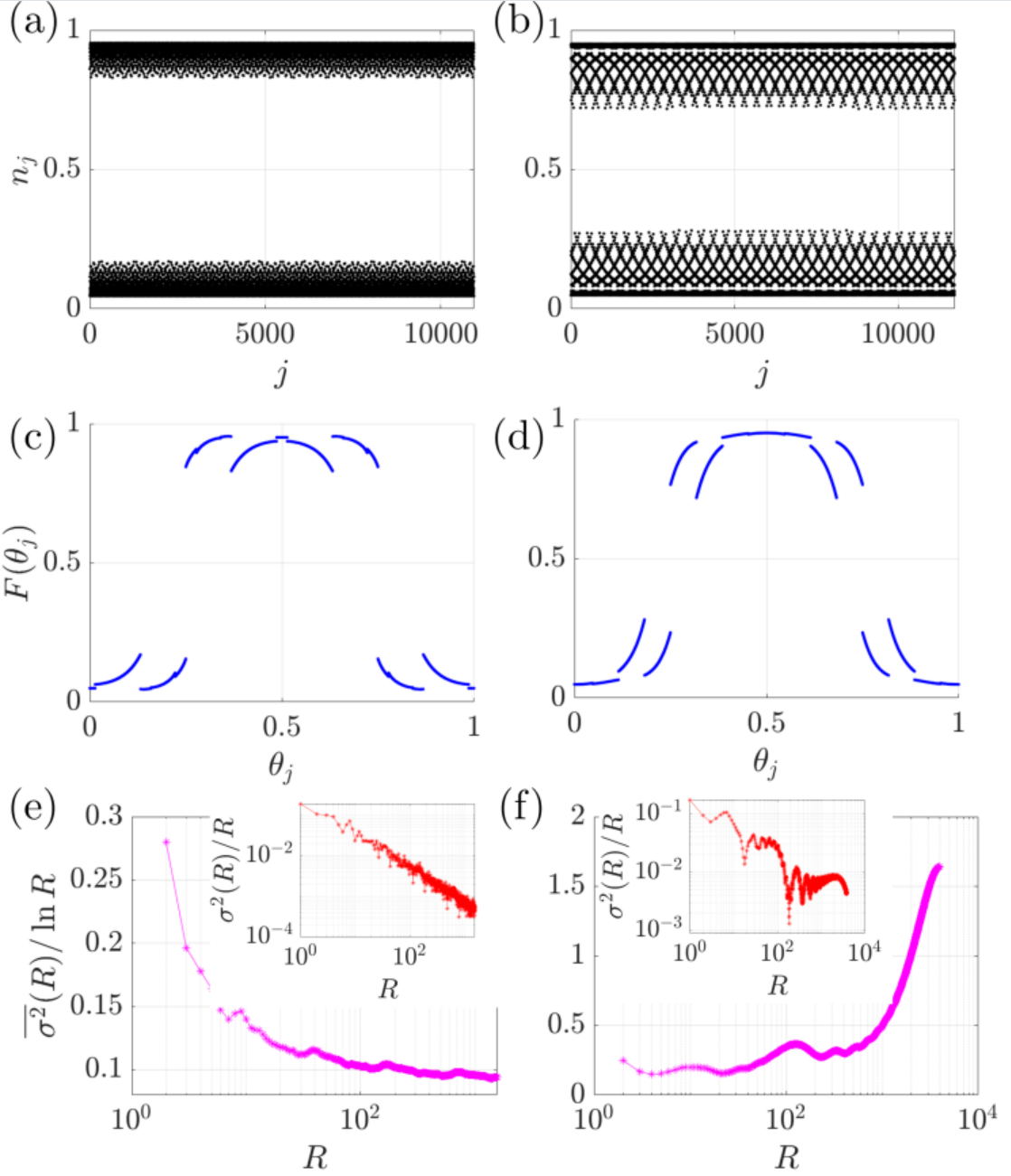}
    \caption{(a,b) Spatial distribution of the local density $n_j$ in the Aubry--Andr\'e model [Eq.~\eqref{eq:AA}], calculated for $\lambda=3t$, $E_F=0$ and different $\beta$'s. (c,d) Hull charge profile $F(\theta_j)$, where $\theta_j=j\beta \ \mathrm{mod} \ 1$. Both are piecewisely regular with finitely many jumps. (e,f) $\overline{\sigma^2}(R)/\ln R$ as the function of window radius $R$. Here, $\overline{\sigma^2}(R)=R^{-1}\int_0^{R}\sigma^2(R')dR'$ is the running average of $\sigma^2(R)$. The insets show $\sigma^2(R)/R$ which vanishes for large $R$.  
    (a,c,e) $\beta=6765/10946$ which is continued fraction approximant of golden ratio with $\mu=2$. The system size is 10946. (b,d,f) $\beta=28517/11724=[2;2,3,5,10,31]$ with $\mu\approx2.5794$. The system size is 11724.
    }
    \label{fig: 1}
\end{figure}

Figure~\ref{fig: 1} exhibits the charge distribution in the  localized regime ($\lambda>2t$) of the Aubry--Andr\'e model for two different irrationals given by continued fractions, $\beta_1=[0;1,1,1,\cdots]=\varphi$ and $\beta_2=[2;2,3,5,10,31,\cdots]$~\cite{Khinchin1997,SM}.
Calculations were done with rational approximants of $6765/10946 (\simeq \beta_1)$ and $28517/11724 (\simeq \beta_2)$, respectively, under periodic boundary conditions.
Note that $\mu(\beta_1)=2$ as the golden ratio is algebraic, while $\mu(\beta_2)=5/2$. Thus, for a piecewise regular $F$ with finitely many jump [see Figs.~\ref{fig: 1}(c,d)], we have $\alpha=1 \ (1/3)$ [i.e., class II (III)] for $\beta_1 \ (\beta_2)$ 
[see Figs.~\ref{fig: 1}(e,f)]. 

\tcm{\textit{Stable gaps and Fermi-level dominance.--}}
Our arithmetic theorem determines how a hull singularity is converted into the infrared fluctuations. A separate result identifies which occupied states contribute such a singularity. Let $E_g (<E_F)$ lie in a spectral gap that remains open for every real hull phase and under an analytic continuation $\theta\mapsto\theta+\mathrm{i}\eta$ throughout a strip $|\eta|<\rho_0$. Decompose the Fermi projector into a near-Fermi shell and a deep occupied sector, $P_F=P_{E_g\le E\le E_F}+P_{E<E_g}$.
Here, $P_{E<E_g}$ projects onto the spectral sector below $E_g$ and the other term denotes the complementary occupied shell. Their local densities satisfy $n_j=n_j^{\rm shell}+n_j^{\rm deep}$, with $n_j^{\rm deep}:=\langle j|P_{E<E_g}|j\rangle$.

Choose a positively oriented closed contour $\Gamma$ in the complex energy plane that encloses the spectral sector below $E_g$ and no other spectrum, and that remains in the resolvent set of $H(\theta)$ uniformly for all $\theta$ in a smaller complex strip $|\operatorname{Im}\theta|\le\rho<\rho_0$. The deep projector is represented by
\begin{equation}
 P_{E<E_g}(\theta)=\frac{1}{2\pi i}\oint_\Gamma dz\,[z-H(\theta)]^{-1}.
 \label{eq:Riesz}
\end{equation}
The possibility of choosing this single uniform contour is precisely where the stable-gap assumption enters. This condition is natural for hull-uniform spectral gaps of analytic quasiperiodic Hamiltonians including generalized Aubry--Andr\'e model, which on the real hull are associated with the gap labels of the gap-labeling theorem~\cite{BellissardBovierGhez1992,BellissardHerrmannZarrouati2000}. The Riesz representation makes the deep hull contribution analytic whose Fourier coefficients decay exponentially. Thus, stable gapped phases universally shows class-I hyperuniform charge distribution. Moreover, the infrared weight and cross term with this contribution of stable gapped sector vanish faster than any power. As a consequence, nontrivial power-law scaling is governed by near-Fermi states, despite that every occupied state equally contributes to the local density.

The stability qualification is essential. A gap that exists only at one selected phase is not a stable hull gap, and a contour satisfying the uniform condition above need not exist. Hence, such a phase-specific gapped system would evade the class-I conclusion without contradicting our theorem. We indeed find that a class-II or class-III gapped phase is possible in phase-constrained particle-hole-symmetric Aubry-Andr\'e model~\cite{SM}.

\tcm{\textit{Discussion and conclusion--}}
Nothing in the arithmetic mechanism is specific to the cosine potential of Eq.~\eqref{eq:AA}. The essential ingredients are translation covariance, a hull representation of the observable~\cite{BellissardHerrmannZarrouati2000}, and control of its large-harmonic Fourier tail. The same framework applies to the generalized Aubry--Andr\'e model~\cite{GaneshanPixleyDasSarma2015} and more general torus-generated quasiperiodic Hamiltonians in which lattice translations act as rotations of the hull phase~\cite{PhysRevLett.62.2714}, whenever the large-harmonic Fourier tail of the relevant hull observable can be controlled. More broadly, Sturmian quasiperiodic chains admit analogous phason and cut-and-project descriptions~\cite{KrausZilberberg2012,Jagannathan2021,BjorklundHartnick2024,PhysRevLett.50.1870}, allowing the same arithmetic mechanism to be applied.

The one-frequency restriction is also not fundamental. For a $d$-frequency hull with frequency vector $\bm\beta$, the scalar harmonic index $m$ is replaced by an integer vector $\bm m\in\mathbb Z^d$, and the relevant infrared scale is set by $\normZ{\bm m\cdot\bm\beta}$. Accordingly, the one-frequency factor $\mu(\beta)-1$ is replaced by the dual resonance exponent $\omega_*(\bm\beta)$, yielding $\alpha+1=\frac{2r}{\omega_*(\bm\beta)}$. The corresponding multi-frequency resonance theorem and stable-gap extension are derived in the Supplemental Material~\cite{SM}.

Our result reveals a distinct arithmetic control of long-wavelength charge fluctuations in quasiperiodic electron systems, beyond its previously established roles in critical scaling and fluctuations of quasiperiodic point distributions~\cite{PhysRevB.98.134201,BjorklundHartnick2024}. Near-Fermi states determine the available high-order hull harmonics, while the irrationality exponent controls their transfer to the infrared. States separated by a stable gap contribute only a smooth background. Thus, the hyperuniformity class emerges from the interplay between near-Fermi physics and the arithmetic of quasiperiodicity. 

\begin{acknowledgments}
 This work was supported by JSPS KAKENHI Grant No.~JP25H01397, JP25H01398, JP25K24854, and JP25K24855.
\end{acknowledgments}

\bibliography{reference}
\nocite{Weyl1916,Kato1995,KuipersNiederreiter1974}

\newpage
\begin{widetext}
\setcounter{equation}{0}
\setcounter{figure}{0}
\renewcommand{\thefigure}{S\arabic{figure}}
\renewcommand{\theequation}{S\arabic{equation}}

\begin{center}
{\large\bfseries Supplemental Material for\\[2mm]
``Irrationality Measure Controls Long-Wavelength Charge Fluctuations\\
in Quasiperiodic Systems''}\\[4mm]
Junmo Jeon and Shiro Sakai\\[1mm]
\textit{Physics Division, Sophia University, Chiyoda-ku, Tokyo 102-8554, Japan}\\[1mm]
\today
\end{center}
\vspace{3mm}

This Supplemental Material collects the mathematical statements behind the Letter and separates them from the physical interpretation emphasized in the main text. We first summarize the real-space diagnostics used for numerical classification and connect them to the reciprocal-space formulation, then prove the exceptional-resonance scaling for a one-frequency hull. We next discuss continued fractions and finite-size arithmetic design, establish the stable-gap analyticity theorem and its scope. Lastly, we generalize the theorem to a multi-frequency torus hull.

\section{Hyperuniformity in real and reciprocal space}

We use the real-space formulation primarily as a numerical diagnostic of hyperuniformity, which can also be defined through long-wavelength fluctuations in momentum space~\cite{TorquatoStillinger2003,Torquato2018}. Consider a one-dimensional scalar field $n_j$ on the integer lattice. For a window of radius $R$ centered at a continuous (or discrete) position $x_c$, define
\begin{equation}
 N_R(x_c):=\sum_{j\in\Z}n_j\,\Theta(R-|j-x_c|),
 \label{S:eq:windowCharge}
\end{equation}
where $\Theta$ is the Heaviside step function. The variance is
\begin{equation}
 \sigma^2(R):=\overline{N_R(x_c)^2}-\overline{N_R(x_c)}^{\,2}.
 \label{S:eq:windowVariance}
\end{equation}
Here, the overline denotes an average over $x_c$. For a Poisson-like distribution in one dimension, $\sigma^2(R)$ is proportional to $R$, i.e., the volume law. It is therefore convenient to define $A(R):=\sigma^2(R)/R$ to identify hyperuniform distribution. Specifically, the distribution is hyperuniform when $\lim_{R\to\infty}A(R)=0$.

Ordered and quasiperiodic structures generally show bounded oscillations as the window size $R$ is varied. Following the real-space diagnostic used in previous studies of quasiperiodic charge density~\cite{SakaiAritaOhtsuki2022,XiangJeonSakai2026}, we smooth these oscillations by averaging over window sizes,
\begin{equation}
 \overline B(R):=\frac{1}{R}\int_0^R\sigma^2(R')\,dR'.
 \label{S:eq:Bbar}
\end{equation}
For a hyperuniform distribution in one dimension, the conventional classes are characterized by different large-$R$ scalings of $\overline B(R)$. In detail, $\overline B(R)$ approaches a constant in class I, grows as $\ln R$ in class II, and grows as $R^{1-\alpha}$ for class III with $0<\alpha<1$. Hence, a particularly useful numerical quantity is
\begin{equation}
 D_B(R):=\frac{\overline B(R)}{\ln R}.
 \label{S:eq:DB}
\end{equation}
Note that for large $R\gg1$, $D_B(R)$ vanishes for class I, approaches a positive constant for class II, and increases without bound for class III. In practice, $A(R)$ and $D_B(R)$ are used together: vanishing $A(R)$ first establishes hyperuniformity, while $D_B(R)$ specifies its class. If $A(R)$ instead approaches a nonzero value, the distribution is nonhyperuniform. Infrared fluctuations stronger than this boundary, i.e., $\lim_{R\to\infty}A(R)\to\infty$ are commonly termed anti-hyperuniform.

The reciprocal-space description follows from the Fourier amplitude of $n_j-\bar n$, where $\bar n$ is the spatial mean. For a segment containing $L$ sites, let $\delta n_L(k):=\sum_{j=1}^{L}(n_j-\bar n)e^{-ikj}$. The finite-size structure factor is
\begin{equation}
 S_L(k):=\frac{|\delta n_L(k)|^2}{L}.
 \label{S:eq:structureFactor}
\end{equation}
Note that the thermodynamic diffraction measure $S(k)$ is obtained as the weak limit of $S_L(k)\,dk$. The Fourier transform of an interval window, $w_R(x)=\Theta(R-|x|)$, is
\begin{equation}
 \widetilde w_R(k)=\frac{2\sin(kR)}{k}.
 \label{S:eq:windowFT}
\end{equation}
Up to the conventional normalization of the structure factor, the center-averaged variance is the spectral weight filtered by this window form factor,
\begin{equation}
 \sigma^2(R)=\int_{\mathbb R}\frac{dk}{2\pi}\,S(k)|\widetilde w_R(k)|^2.
 \label{S:eq:varianceSpectral}
\end{equation}
This relation gives the standard equivalence between the large-$R$ and small-$k$ classifications. If $S(k)\sim |k|^\alpha$, then class I corresponds to $\alpha>1$, class II to $\alpha=1$, and class III to $0<\alpha<1$. At $\alpha=0$ the structure factor no longer vanishes in the infrared, while $\alpha<0$ describes anti-hyperuniform enhancement.

For a quasiperiodic pure-point spectrum, it is convenient to work with the integrated low-wave-number weight~\cite{PhysRevB.95.054119}
\begin{equation}
 Z(K):=\int_{|k|<K}S(k)\,dk.
 \label{S:eq:generalZ}
\end{equation}
An ordinary power law $S(k)\sim |k|^\alpha$ gives $Z(K)\sim K^{\alpha+1}$. The proof below is formulated directly in terms of $Z(K)$ because the Bragg spectrum of charge density need not be represented by a smooth function in general. The real-space quantities $A(R)$, $\overline B(R)$, and $D_B(R)$ are then natural diagnostics for numerical support of the resulting infrared class.

\section{Translation-covariant hull representation}

We work on the one-dimensional lattice Hilbert space $\ell^2(\Z)$ with orthonormal site basis $\{\vert j\rangle:j\in\Z\}$. The one-site translation operator $T$ is defined by $T\vert j\rangle=|j+1\rangle$. Let $\theta\in\mathbb R/(2\pi\Z)$ be a $2\pi$-periodic hull phase and let $\beta\notin\mathbb Q$ be the modulation frequency (see Fig.~\ref{fig: S_hull}). A one-frequency translation-covariant Hamiltonian~\cite{BellissardHerrmannZarrouati2000} is a family $H(\theta)$ satisfying
\begin{equation}
 T H(\theta)T^{-1}=H(\theta-2\pi\beta).
 \label{S:eq:cov}
\end{equation}
If $P(\theta)$ is a spectral operator constructed covariantly from $H(\theta)$, such as a Fermi projector or a spectral-shell projector, then it obeys the same covariance,
\begin{equation}
 T P(\theta)T^{-1}=P(\theta-2\pi\beta).
 \label{S:eq:Pcov}
\end{equation}
The local density is given by
\begin{equation}
 n_j(\theta):=\langle j|P(\theta)\vert j\rangle.
\end{equation}
Since $\vert j\rangle=T^j\vert0\rangle$, using Eq.~\eqref{S:eq:Pcov},
\begin{align}
 n_j(\theta)
 &=\langle0|P(\theta+2\pi\beta j)|0\rangle\\
 &:=F(\theta+2\pi\beta j),
 \label{S:eq:hullrep}
\end{align}
where $F(\vartheta):=\langle0|P(\vartheta)|0\rangle$ is the corresponding scalar observable on the hull.
\begin{figure}[h]
    \centering
    \includegraphics[width=0.5\textwidth]{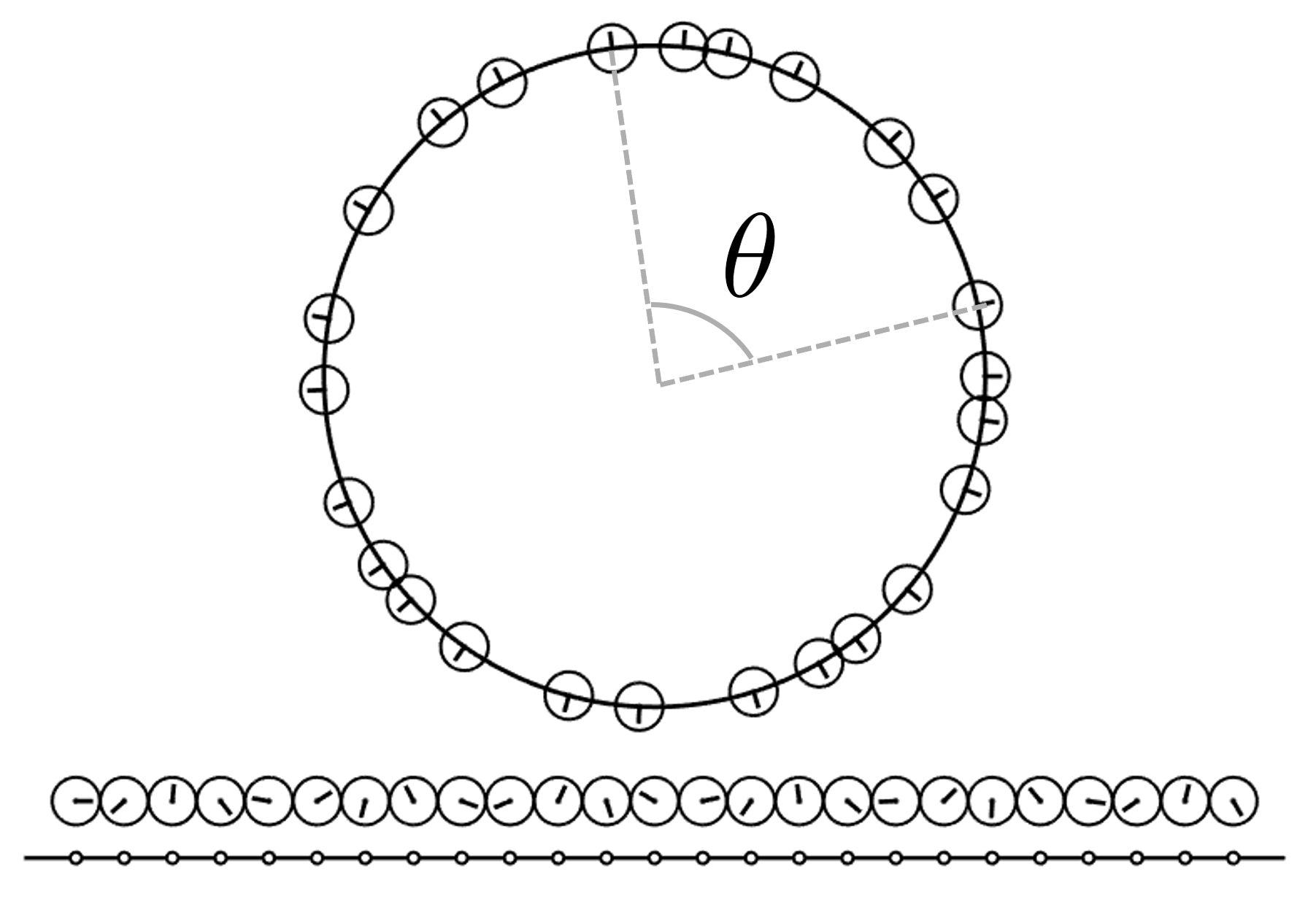}
    \caption{Translation-covariant hull representation. Spatial translation operator corresponds to rotation operator of hull phase $\theta\in\mathbb R/(2\pi\Z)$. For irrational modulation frequency $\beta$, hull phase densely covers a circle, and hence the hull of quasiperiodic chain is a circle, $S^1=\mathbb R/(2\pi\Z)$.
    }
    \label{fig: S_hull}
\end{figure}

Let
\begin{equation}
 \overline F:=\frac{1}{2\pi}\int_0^{2\pi}F(\vartheta)\,d\vartheta
\end{equation}
be the hull average, and define the Fourier coefficients
\begin{equation}
 c_m:=\frac{1}{2\pi}\int_0^{2\pi}
 [F(\vartheta)-\overline F]e^{-im\vartheta}\,d\vartheta,
 \qquad m\in\Z\setminus\{0\}.
\end{equation}
Then
\begin{equation}
 F(\vartheta)-\overline F=\sum_{m\ne0}c_m e^{im\vartheta},
 \label{S:eq:Fourier}
\end{equation}
and sampling along the physical lattice gives
\begin{equation}
 n_j(\theta)-\overline F=
 \sum_{m\ne0}c_m e^{im\theta}e^{i2\pi m\beta j}.
 \label{S:eq:sampledFourier}
\end{equation}
Each hull harmonic therefore produces a Bragg peak at
$k=2\pi m\beta$ modulo the reciprocal lattice. In the thermodynamic
diffraction measure, distinct harmonics are orthogonal for irrational
$\beta$. Therefore, the Bragg intensity associated with the $m$-th harmonic is
proportional to $|c_m|^2$.

For a real number $x$, we write
\begin{equation}
 \normZ{x}:=\min_{\ell\in\Z}|x-\ell|
 \label{S:eq:normZdef}
\end{equation}
for its distance to the nearest integer. With lattice spacing set to unity, the magnitude of the $m$-th physical Bragg wave number modulo the reciprocal lattice is therefore
\begin{equation}
 k_m=2\pi\normZ{m\beta}.
 \label{S:eq:km}
\end{equation}
For scaling we remove the inessential factor $2\pi$ from the wave-number cutoff. The general integrated weight in Eq.~\eqref{S:eq:generalZ} then specializes to
\begin{equation}
 Z(K)=\sum_{m\ne0:\,\normZ{m\beta}<K}|c_m|^2
 \label{S:eq:Z}
\end{equation}
up to a normalization that does not affect logarithmic exponents.
For a pure-point quasiperiodic diffraction spectrum, this sum is precisely the total amount of long-wavelength Bragg weight relevant to the theorem below.

\section{Irrationality exponent and exceptional resonances}

For $\beta\notin\mathbb Q$, the standard irrationality exponent~\cite{Bugeaud2004} is
\begin{equation}
 \mu(\beta):=\sup\left\{\nu>0:
 \left|\beta-\frac pq\right|<q^{-\nu}
 \text{ for infinitely many }(p,q)\in\Z\times\N\right\}.
 \label{S:eq:mu}
\end{equation}
For each natural number $q$, let $p$ be the nearest integer to $q\beta$. Then
\begin{equation}
 \left|\beta-\frac pq\right|=\frac{\normZ{q\beta}}{q}.
 \label{S:eq:muomegaIdentity}
\end{equation}
This identity shifts the approximation exponent by exactly one. Defining the one-frequency resonance exponent
\begin{equation}
 \omega(\beta):=\limsup_{m\to\infty}
 \frac{-\log\normZ{m\beta}}{\log m},
 \label{S:eq:omega}
\end{equation}
we obtain
\begin{equation}
 \omega(\beta)=\mu(\beta)-1.
 \label{S:eq:omegaMu}
\end{equation}
Indeed, an approximation error $|\beta-p/q|\sim q^{-\mu}$ corresponds to a phase mismatch $\normZ{q\beta}\sim q^{-(\mu-1)}$ on the same logarithmic subsequence.

For the remainder of the one-frequency proof, we write $\omega:=\omega(\beta)$ and assume $\omega<\infty$. The definition of the limsup gives two complementary statements. For every fixed $\epsilon>0$, there exists an integer $M_\epsilon$ such that
\begin{equation}
 \normZ{m\beta}\ge m^{-(\omega+\epsilon)}
 \qquad {\rm for}\quad ^\forall m\ge M_\epsilon,
 \label{S:eq:eventualLower}
\end{equation}
whereas for every $\epsilon>0$ there are infinitely many positive integers $m$ satisfying
\begin{equation}
 \normZ{m\beta}<m^{-(\omega-\epsilon)}.
 \label{S:eq:infApprox}
\end{equation}
We emphasize that no algebraicity assumption enters these statements.

\subsection{Main theorem}
Now let us formally state our main theorem.
Assume that there exists a constant $B>0$, an exponent $r>1/2$, and an index $m_0$ such that the global upper bound
\begin{equation}
 |c_m|\le B|m|^{-r}
 \label{S:eq:tailUpper}
\end{equation}
holds for all $|m|\ge m_0$. We also impose a nondegeneracy condition only on the exceptional
resonances. Let $\{q_j\}$ be a sequence of positive integers such that $\frac{-\log\normZ{q_j\beta}}{\log q_j}\to \omega(\beta)$ with
$q_j\to\infty$. We refer to any sequence satisfying this condition as an \emph{exponent-realizing sequence}. We assume that there exist a constant $A>0$ and an exponent-realizing sequence $\{q_j\}$ for which
\begin{equation}
 |c_{q_j}|\ge A q_j^{-r}.
 \label{S:eq:tailLower}
\end{equation}
Note that no lower bound assumption is required for generic large Fourier indices. The second condition only prevents the exponent-realizing arithmetic resonances from being removed by zeros or cancellations of the observable's Fourier coefficients. For the logarithmic scaling below, Eq.~\eqref{S:eq:tailLower} can be weakened to require only the same logarithmic decay exponent $r$ along an exponent-realizing subsequence (see Sec.~\ref{S:subsec:logNondegeneracy} for details.). The global upper bound in Eq.~\eqref{S:eq:tailUpper} gives
$\alpha+1\ge 2r/[\mu(\beta)-1]$, whereas the nondegeneracy condition provides the matching reverse inequality. The condition is relevant to algebraic Fourier tails of gapless charge profiles; stable-gap sectors are treated separately below and give exponentially decaying Fourier coefficients. We discuss
the physical meaning and robustness of the nondegeneracy condition for gapless systems after establishing the following exceptional-resonance theorem.

\textbf{Theorem S1.} If $2\le\mu(\beta)<\infty$, then
\begin{equation}
 \liminf_{K\to0^+}\frac{\log Z(K)}{\log K}
 =\frac{2r}{\mu(\beta)-1}.
 \label{S:eq:theorem}
\end{equation}
Since $\omega=\mu(\beta)-1$, the right-hand side is $2r/\omega$. If the singular-envelope structure-factor exponent $\alpha$ is defined by
\begin{equation}
 \alpha+1:=\liminf_{K\to0^+}\frac{\log Z(K)}{\log K},
 \label{S:eq:alphadef}
\end{equation}
then
\begin{equation}
 \alpha=\frac{2r}{\mu(\beta)-1}-1.
 \label{S:eq:alpha}
\end{equation}
We prove Eq.~\eqref{S:eq:theorem} by showing 
that the lower and upper bounds of the left-hand side agree.

\subsection{Proof for lower bound}
For a fixed $K$, let
\begin{equation}
 \mathcal R_K^+:=\{m\in\N:\normZ{m\beta}<K\}
 =\{m_1<m_2<m_3<\cdots\}
 \label{S:eq:resonantSet}
\end{equation}
be the ordered set of positive $K$-resonant Fourier indices. Now for a fixed $\epsilon>0$, let us choose $M_\epsilon$ as in Eq.~\eqref{S:eq:eventualLower}. Since $\beta$ is irrational, the finite set $1\le m<M_\epsilon$ has a strictly positive minimum
\begin{equation}
 \delta_\epsilon:=\min_{1\le m<M_\epsilon}\normZ{m\beta}>0.
 \label{S:eq:deltaepsilon}
\end{equation}
Take $K<\delta_\epsilon/2$. If $m_j>m_\ell$ are two elements of $\mathcal R_K^+$, the triangle inequality for the distance modulo integers gives
\begin{align}
 \normZ{(m_j-m_\ell)\beta}
 &\le \normZ{m_j\beta}+\normZ{m_\ell\beta}<2K<\delta_\epsilon.
 \label{S:eq:resonanceDifference}
\end{align}
The difference $d:=m_j-m_\ell$ cannot be smaller than $M_\epsilon$; otherwise Eq.~\eqref{S:eq:deltaepsilon} would be violated. Thus, $d\geq M_\epsilon$ and hence Eq.~\eqref{S:eq:eventualLower} implies
\begin{equation}
 d^{-(\omega+\epsilon)}\leq \normZ{d\beta}<2K,
\end{equation}
or
\begin{equation}
 m_j-m_\ell>(2K)^{-1/(\omega+\epsilon)}.
 \label{S:eq:spacing}
\end{equation}
The same reasoning applied to the first resonance $m_1$, after decreasing the allowed $K$ if necessary. In detail, for a sufficiently small $K$, finite number of $m<M_\epsilon$ are excluded from $R_K^+$, and hence $m_1^{-(\omega+\epsilon)}\le\normZ{m_1\beta}<K$ for $m_1\ge M_\epsilon$. This yields a constant $C_\epsilon>0$ such that
\begin{equation}
 m_j\ge C_\epsilon j K^{-1/(\omega+\epsilon)}.
 \label{S:eq:sep}
\end{equation}
Using Eqs.~\eqref{S:eq:Z} and \eqref{S:eq:tailUpper} and including negative Fourier indices by a factor of two, we obtain
\begin{align}
 Z(K)
 &\le 2B^2\sum_{j\ge1}m_j^{-2r}\\
 &\le \tilde{C_\epsilon} K^{2r/(\omega+\epsilon)}
 \sum_{j\ge1}j^{-2r}.
 \label{S:eq:Zupper}
\end{align}
Here, $\tilde{C_\epsilon}>0$ is an inessential coefficient. The series converges because $r>1/2$. Since $\log K<0$ for small $K<1$, Eq.~\eqref{S:eq:Zupper} gives
\begin{equation}
 \liminf_{K\to0^+}\frac{\log Z(K)}{\log K}
 \ge\frac{2r}{\omega+\epsilon}.
\end{equation}
Taking $\epsilon\downarrow0$ gives
\begin{equation}
 \liminf_{K\to0^+}\frac{\log Z(K)}{\log K}
 \ge\frac{2r}{\omega}.
 \label{S:eq:upperResult}
\end{equation}

\subsection{Proof for upper bound}

Let $\{q_j\}$ be the exponent-realizing sequence entering the
nondegeneracy condition in Eq.~\eqref{S:eq:tailLower}, and define
\begin{equation}
 a_j:=-\frac{\log\normZ{q_j\beta}}{\log q_j}.
 \label{S:eq:aj}
\end{equation}
By the definition of an exponent-realizing sequence, $a_j$ approaches $\omega$ as $j$ goes to infinity. We choose the corresponding infrared cutoffs as
\begin{equation}
 K_j:=2\normZ{q_j\beta}.
 \label{S:eq:Kj}
\end{equation}
The factor of two ensures
$\normZ{q_j\beta}<K_j$, so that the $q_j$-th harmonic is included in the
sum defining $Z(K_j)$. The nondegeneracy condition
Eq.~\eqref{S:eq:tailLower} gives the single-mode lower bound
\begin{equation}
 Z(K_j)\ge |c_{q_j}|^2\ge A^2q_j^{-2r}.
 \label{S:eq:singleModeLower}
\end{equation}
From Eq.~\eqref{S:eq:aj}, we have
\begin{equation}
 \normZ{q_j\beta}=q_j^{-a_j},
\end{equation}
and hence
\begin{equation}
 \log K_j=\log 2-a_j\log q_j.
 \label{S:eq:logKj}
\end{equation}
Since $a_j\to\omega>0$ as $q_j\to\infty$, one has
$K_j\to0$. Thus, for sufficiently large $j$, $\log K_j<0$. Taking the
logarithm of Eq.~\eqref{S:eq:singleModeLower} gives
\begin{equation}
 \log Z(K_j)\ge 2\log A-2r\log q_j.
\end{equation}
For small $K_j<1$, division by $\log K_j$ reverses the inequality,
so that
\begin{align}
 \frac{\log Z(K_j)}{\log K_j}
 &\le
 \frac{2\log A-2r\log q_j}
 {\log 2-a_j\log q_j}
 \nonumber\\
 &=
 \frac{
 2r-\dfrac{2\log A}{\log q_j}
 }{
 a_j-\dfrac{\log 2}{\log q_j}
 }.
 \label{S:eq:subsequenceExponent}
\end{align}
As $a_j\to\omega$ and $\log q_j\to\infty$, the right-hand side
converges to $2r/\omega$. Therefore
\begin{equation}
 \limsup_{j\to\infty}
 \frac{\log Z(K_j)}{\log K_j}
 \le \frac{2r}{\omega}.
 \label{S:eq:subsequenceUpper}
\end{equation}

Finally, $\{K_j\}$ is a sequence tending to zero, whereas the
definition of the logarithmic exponent takes the lower limit over
all $K\to0^+$. The lower limit over all infrared scales cannot exceed
the lower limit along any particular sequence of infrared scales.
Consequently,
\begin{align}
 \liminf_{K\to0^+}\frac{\log Z(K)}{\log K}
 &\le
 \liminf_{j\to\infty}
 \frac{\log Z(K_j)}{\log K_j}
 \nonumber\\
 &\le
 \limsup_{j\to\infty}
 \frac{\log Z(K_j)}{\log K_j}
 \le
 \frac{2r}{\omega}.
 \label{S:eq:lowerResult}
\end{align}
Combining Eq.~\eqref{S:eq:lowerResult} with
Eq.~\eqref{S:eq:upperResult} yields
\begin{equation}
 \liminf_{K\to0^+}\frac{\log Z(K)}{\log K}
 =\frac{2r}{\omega},
\end{equation}
which proves Theorem~S1.

\section{Physical interpretation of the exponents}
\subsection{Hyperuniformity class and scaling exponent $\alpha$}
The scaling exponent $\alpha$ in Eq.~\eqref{S:eq:alphadef} classifies long-wavelength behavior of the structure factor in terms of hyperuniformity. Although we implicitly assume $Z(K)>0$ for every nonzero $K$ under consideration in Eq.~\eqref{S:eq:alphadef}, we may consider the real-extension of logarithmic function to include a special type of hyperuniformity known as stealthy hyperuniform. For systems with dense Bragg spectra, including quasiperiodic systems, $S(k)$ is a pure-point measure, so the pointwise limit $S(k)\to0$ is generally ill-defined. We therefore characterize the long-wavelength density fluctuations in terms of the integrated structure factor $Z(K)$. Specifically, hyperuniform, non-hyperuniform, and antihyperuniform behavior correspond, respectively, to
\begin{equation}
\lim_{K\to0^+}\frac{Z(K)}{K}
=
\begin{cases}
0, & \text{hyperuniform},\\
c\in(0,\infty), & \text{non-hyperuniform},\\
\infty, & \text{antihyperuniform}.
\end{cases}
\label{S:eq:HUcriterion}
\end{equation}
Then, the scaling exponent $\alpha$ determines whether Eq.~\eqref{S:eq:HUcriterion} holds whenever $\alpha\neq0$, whereas $\alpha=0$ is a marginal case for which the algebraic exponent alone is insufficient.

\begin{table}[t]
\caption{\label{S:tab:alpha_classification}
Classification according to the scaling exponent $\alpha$ and the corresponding relation between the regularity exponent $r$ and the resonance exponent $\omega$, following $\alpha+1=2r/\omega$. The marginal case $\alpha=0$ requires additional information on the subalgebraic behavior of $Z(K)$.}
\begin{ruledtabular}
\begin{tabular}{cccc}
Scaling exponent & Relation between $r$ and $\omega$ & $Z(K)/K$ & Hyperuniformity classification \tabularnewline
\colrule
$\alpha>1$ & $r>\omega$ & $\to0$ & class I\tabularnewline
$\alpha=1$ & $r=\omega$ & $\to0$ & class II\tabularnewline
$0<\alpha<1$ & $\omega/2<r<\omega$ & $\to0$ & class III \tabularnewline
$\alpha=0$ & $r=\omega/2$ & Undetermined & Inconclusive \tabularnewline
$\alpha<0$ & $1/2<r<\omega/2$ & $\to\infty$ & Anti-hyperuniform \tabularnewline
\end{tabular}
\end{ruledtabular}
\end{table}

For $\alpha>0$, choose any $0<\delta<\alpha$. Then, for all sufficiently small $K$,
\begin{equation}
\frac{\log Z(K)}{\log K}\geq 1+\alpha>1+\delta.
\end{equation}
Since $\log K<0$, $\frac{Z(K)}{K}<K^\delta\longrightarrow0$.
Thus, any positive value of $\alpha$ guarantees hyperuniformity.

Similarly, for $\alpha<0$, choose $0<\delta<-\alpha$. Then, there exists a sequence $K_j\to0$ such that
\begin{equation}
\frac{\log Z(K_j)}{\log K_j}=1+\alpha <1-\delta.
\end{equation}
Since $\log K_j<0$, this gives $\frac{Z(K_j)}{K_j}>K_j^{-\delta}\longrightarrow\infty$.
Thus, $\alpha<0$ necessarily implies anti-hyperuniformity.

On the other hand, the marginal value $\alpha=0$ is inconclusive since Eq.~\eqref{S:eq:alphadef} detects the leading algebraic exponent but is insensitive to logarithmic, or more generally subalgebraic, corrections. Consequently, systems with distinct long-wavelength fluctuation properties can share the same value $\alpha=0$. Representative examples are summarized in Table~\ref{S:tab:alpha_zero}.

\begin{table}[t]
\caption{\label{S:tab:alpha_zero}
Distinct small-$K$ behaviors sharing the same $\alpha=0$. 
}
\begin{ruledtabular}
\begin{tabular}{cccc}
$Z(K)$ & $Z(K)/K$ & $I_n$, $S(k)=\sum_{\vert n\vert\ge n_0>0}^{\infty}I_n\delta(k-1/n)$ & Behavior \tabularnewline
\colrule
$\displaystyle \frac{K}{\log (1/K)}$
& $\displaystyle \frac{1}{\log(1/K)}\to0$
& $\displaystyle \frac{1}{n^2 \log n}$
& Hyperuniform \tabularnewline
$cK$, $c>0$
& $c$
& $\displaystyle \frac{c}{n^2}$
& Non-hyperuniform \tabularnewline
$K\log(1/K)$
& $\log(1/K)\to\infty$
& $\displaystyle \frac{\log n}{n^2}$
& Anti-hyperuniform \tabularnewline
\end{tabular}
\end{ruledtabular}
\end{table}

Indeed, each of the three examples in Table~\ref{S:tab:alpha_zero} satisfies
\begin{equation}
\lim_{K\to0^+}
\frac{\log Z(K)}{\log K}=1,
\end{equation}
and therefore gives $\alpha=0$. Nevertheless, their ratios $Z(K)/K$ respectively vanish, remain finite and nonzero, or diverge. Thus, $\alpha=0$ should not by itself be identified with non-hyperuniformity. Hence, a definite classification at $\alpha=0$ requires the ratio in Eq.~\eqref{S:eq:HUcriterion}.

Additionally, although the preceding discussion implicitly assumes $Z(K)>0$ for every nonzero $K$ under consideration, Eq.~\eqref{S:eq:alphadef} admits a natural extension to stealthy hyperuniform, which has vanishing structure factor in a finite neighborhood of zero momentum, and hence there exists $K_c>0$ such that
\begin{equation}
Z(K)=0,
\qquad
0<K<K_c.
\end{equation}
The ordinary logarithm in Eq.~\eqref{S:eq:alphadef} is not defined at $Z(K)=0$. However, one can naturally extend the logarithmic funtion by adopting the extended-real convention $\log 0:=-\infty$. Since $-\infty<\log K<0$ for sufficiently small $K\neq0$, this natural extension gives
\begin{equation}
\frac{\log Z(K)}{\log K}=\infty,
\qquad
0<K<K_*,
\end{equation}
and hence $\alpha=\infty$.
Thus, stealthy hyperuniformity is naturally incorporated as an infinitely strong endpoint of the hyperuniform side of the algebraic scaling classification. Note that the converse does not generally hold since $\alpha=+\infty$ only means that $Z(K)$ vanishes faster than every finite algebraic power. For example, $Z(K)\sim e^{-1/K}$ remains strictly positive for every $K>0$ but $\lim_{K\to0^+}
\frac{\log Z(K)}{\log K}=\infty$.
Hence stealthy hyperuniformity is a special case of the $\alpha=\infty$ endpoint, distinguished by the exact vanishing of $Z(K)$ over a finite interval. 

\textbf{Theorem~S1} establishes that the scaling exponent $\alpha$ governing the long-wavelength density fluctuations is determined by the interplay between the regularity exponent $r$ and the irrationality exponent $\mu$ (or equivalently, $\omega$). Table~\ref{S:tab:alpha_classification} gives the corresponding relations between $r$ and $\omega$ and the resulting hyperuniformity classifications. In the following, we explain the physical meanings of these two exponents in detail.

\subsection{Regularity as hull profile and Fourier tails exponent $r$}
The exponent $r$ in \textbf{Theorem~S1} characterizes the algebraic Fourier
tail of the hull observable and should not, in general, be identified
with a differentiability order. Nevertheless, regularity of the hull
profile provides useful sufficient bounds on the Fourier amplitudes.

Let $F\in C^s(\mathbb T)$ for an integer $s\ge1$. Repeated integration
by parts gives
\begin{equation}
 c_m=
 \frac{1}{(im)^s}
 \frac{1}{2\pi}
 \int_0^{2\pi}
 F^{(s)}(\theta)e^{-im\theta}\,d\theta ,
 \label{S:eq:CsFourier}
\end{equation}
and hence
\begin{equation}
 |c_m|
 \le
 \frac{\|F^{(s)}\|_{L^1}}{2\pi}|m|^{-s}.
 \label{S:eq:CsBound}
\end{equation}
The Riemann--Lebesgue lemma gives
$c_m=o(|m|^{-s})$. Hence $C^s$ regularity controls the global
upper tail but does not by itself provide the matching lower bound
required for the equality in Theorem~S1.

Using only the lower bound part of the proof gives
\begin{equation}
 \liminf_{K\to0^+}
 \frac{\log Z(K)}{\log K}
 \ge
 \frac{2s}{\mu(\beta)-1}.
 \label{S:eq:CsInfrared}
\end{equation}
Thus, $s>\mu(\beta)-1$ is a sufficient condition for class-I
hyperuniformity. Note that the equality at a finite algebraic exponent requires
additional information on the actual Fourier tail, including the
nondegeneracy condition on the exponent-realizing resonances.

\subsection{The physical role of the irrationality exponent $\mu$}
The Fourier coefficient $c_m$ determines the weight carried by the $m$-th hull harmonic, whereas the arithmetic of the modulation frequency $\beta$, which characterizes the quasiperiodic structure of the Hamiltonian itself and is independent of the choice of the Fermi level, determines the physical momentum at which that harmonic appears. Hence, the irrationality exponent $\mu$, or equivalently the resonance exponent $\omega=\mu-1$, controls how efficiently high-order hull harmonics are transferred to the physical infrared.

Let $\{q_j\}$ be an exponent-realizing sequence. Along this sequence,
\begin{equation}
 \normZ{q_j\beta}
 =
 q_j^{-\omega+o(1)},
 \qquad
 \omega=\mu(\beta)-1.
\end{equation}
This implies that the $q_j$-th hull harmonic appears at the physical momentum
scale
\begin{equation}
 K_j=q_j^{-\omega+o(1)}.
\end{equation}
If the corresponding Fourier weight has the algebraic scaling
\begin{equation}
 |c_{q_j}|^2=q_j^{-2r+o(1)},
\end{equation}
eliminating $q_j$ gives
\begin{equation}
 |c_{q_j}|^2
 =
 K_j^{\,2r/\omega+o(1)}.
\end{equation}
Thus, $\omega$ controls how strongly that weight is compressed
toward zero physical momentum. The resulting singular-envelope
exponent is
\begin{equation}
 \alpha+1
 =
 \frac{2r}{\omega}
 =
 \frac{2r}{\mu(\beta)-1}.
\end{equation}
As a result, for a fixed hull regularity $r$, increasing $\mu$ shifts the same high-order Fourier weight to smaller momenta and reduces $\alpha$, enhancing long-wavelength charge fluctuations.

This arithmetic mechanism has a direct near-commensurability
interpretation. Let $p/q$ be a rational approximant of $\beta$.
Translation by $q$ lattice sites changes the hull phase by
$2\pi q\beta$. After subtracting the integer winding $2\pi p$, the
residual phase mismatch is
\begin{equation}
 \Delta\varphi_q
 :=
 2\pi\normZ{q\beta}.
 \label{S:eq:phaseMismatch}
\end{equation}
Along an exponent-realizing sequence,
\begin{equation}
 \Delta\varphi_q
 =
 q^{-(\mu-1)+o(1)}.
\end{equation}
The corresponding $q$'s form an exceptional sequence of resonant length scales at which the quasiperiodic pattern returns exceptionally close to the original hull configuration. A larger $\mu$ makes the phase mismatch decay more rapidly along this sequence. Since the phase mismatch determines the physical momentum of the corresponding high-order hull harmonic, a larger $\mu$ also brings the associated Fourier weight to smaller momentum, providing the arithmetic origin of the enhanced long-wavelength charge fluctuations.

\section{Arithmetic classes, continued fractions, and finite-size design}
Write the simple continued fraction of $\beta$ as
\begin{equation}
 \beta=[b_0;b_1,b_2,\ldots],
 \label{S:eq:CFdef}
\end{equation}
where $b_0\in\Z$ and $b_n\in\N$ for $n\ge1$. This notation implies $$\beta=b_0+\frac{1}{b_1+\frac{1}{b_2}+\cdots}.$$ Then, the $n$-th convergent of $\beta$ is the rational number
\begin{equation}
 \frac{p_n}{q_n}:=[b_0;b_1,\ldots,b_n],
 \label{S:eq:convergentDef}
\end{equation}
written in the lowest terms with the denominator $q_n>0$. A standard continued-fraction identity~\cite{Bugeaud2004,Khinchin1997} gives
\begin{equation}
 \mu(\beta)=2+\limsup_{n\to\infty}
 \frac{\log b_{n+1}}{\log q_n}.
 \label{S:eq:CFmu}
\end{equation}
Thus, larger partial quotients $b_{n+1}$ generate better rational approximants.

For the golden-ratio conjugate
\begin{equation}
 \beta_G:=\frac{\sqrt5-1}{2}=[0;1,1,1,\ldots],
\end{equation}
all partial quotients after $b_0$ equal one. Equation~\eqref{S:eq:CFmu} therefore gives
\begin{equation}
 \mu(\beta_G)=2.
\end{equation}
More generally, all algebraic irrational numbers have $\mu=2$~\cite{Roth1955}. 
In particular,
\begin{equation}
 \mu>2\quad\Longrightarrow\quad \beta\ \text{is transcendental}.
\end{equation}
The converse is false: transcendental numbers may also have $\mu=2$. 
\textbf{Theorem~S1} 
therefore distinguishes Diophantine classes rather than providing a complete algebraic/transcendental dichotomy.

A prescribed finite irrationality exponent $\mu_*>2$ can be generated recursively. Once the denominator $q_n$ of the $n$-th convergent is known, choose the $(n+1)$-th partial quotient as
\begin{equation}
 b_{n+1}:=\left\lceil q_n^{\mu_*-2}\right\rceil
 \label{S:eq:construct}
\end{equation}
for all sufficiently large $n$, where $\lceil x\rceil$ denotes the smallest integer not less than $x$. Then, Eq.~\eqref{S:eq:CFmu} gives $\mu(\beta)=\mu_*$. It is also sufficient to insert such large partial quotients only on a sparse subsequence while keeping the remaining $b_n$ bounded.

From the recursion relation of continued fraction, neighboring convergents obey
\begin{equation}
 \vert p_{n+1}q_n-p_nq_{n+1}\vert=1.
 \label{S:eq:detIdentity}
\end{equation}
For a finite system of size $L=q_{n+1}$, choose the modulation frequency
\begin{equation}
 \beta_L:=\frac{p_{n+1}}{L}.
 \label{S:eq:finiteApprox}
\end{equation}
as the $(n+1)$-th approximant of target $\beta\notin\mathbb{Q}$.
Then, Eq.~\eqref{S:eq:detIdentity} gives the exact finite-size resonance
\begin{equation}
 \normZ{q_n\beta_L}=\frac1{q_{n+1}}.
 \label{S:eq:finiteResonance}
\end{equation}
This motivates the finite-size effective exponent
\begin{equation}
 \mu_{L}:=1+\frac{\log q_{n+1}}{\log q_n}.
 \label{S:eq:mueff}
\end{equation}
For a hierarchy designed by Eq.~\eqref{S:eq:construct}, $\mu_{L}$ approaches $\mu_*$ as $n\to\infty$. If the hull tail has the algebraic form
$|c_m|\sim |m|^{-r}$, the resonant mode at $m=q_n$ carries the
weight $|c_{q_n}|^2\sim q_n^{-2r}$ at the infrared scale
$K\sim q_{n+1}^{-1}=q_n^{-(\mu_{L}-1)}$.
Eliminating $q_n$ gives
\begin{equation}
 Z(K)\sim K^{2r/(\mu_{L}-1)},
\end{equation}
and hence
\begin{equation}
 \alpha_{L}\simeq
 \frac{2r}{\mu_{L}-1}-1.
\end{equation}
Thus, finite rational approximants retain the arithmetic scaling
associated with the underlying irrational hierarchy, allowing the
interplay between the hull Fourier-tail exponent and exceptional
arithmetic resonances to be probed directly in finite-size calculations
through the growth of successive convergent denominators.

\section{Nondegeneracy condition and its genericity}

Let us discuss the second assumption of \textbf{Theorem~S1}, namely the requirement that the algebraic Fourier weight remains nonvanishing
along at least one subsequence realizing the arithmetic resonance exponent. We first illustrate its physical meaning in the notable case of $r=1$, and then give a more general argument showing that persistent cancellation is nongeneric for a broad class of hull profiles with finitely many singularities moving with the Fermi energy. Lastly, we show that the fixed-amplitude form of the nondegeneracy condition can be weakened to a logarithmic condition without changing the resulting scaling exponent.

An $r=1$ Fourier tail is naturally produced by isolated jump-like features of a hull charge profile as found in the localized regime (see Fig. 1 of the main text). For the conventional golden-ratio Aubry--Andr\'e model,
$r=1$ and $\mu(\beta)=2$ give $\alpha=1$, in consistency with the class-II behavior observed in the localized regime~\cite{SakaiAritaOhtsuki2022,XiangJeonSakai2026}.

\subsection{Atomic limit and the origin of the $r=1$ tail}

The origin of the $1/|m|$ tail is particularly transparent in the
atomic limit of the Aubry--Andr\'e model. Let $f_j^\dagger$ and $f_j$ respectively 
denote the spinless-fermion creation and annihilation operators at
site $j$. Setting the nearest-neighbor hopping amplitude to $t=0$
gives
\begin{equation}
 H(\theta)
 =
 \lambda\sum_{j\in\mathbb Z}
 \cos(2\pi\beta j+\theta)
 f_j^\dagger f_j ,
 \label{S:eq:atomicH}
\end{equation}
where $\lambda$ is the quasiperiodic modulation strength. At zero
temperature, the hull profile of the local charge density is
\begin{equation}
 F(\theta)
 =
 \Theta(E_F-\lambda\cos\theta),
 \label{S:eq:atomicF}
\end{equation}
where $\Theta$ is the Heaviside step function. For
$|E_F|<|\lambda|$, define
\begin{equation}
 \theta_F:=\arccos(E_F/\lambda).
\end{equation}
The hull profile then has two jump discontinuities within one period,
and its nonzero Fourier coefficients are
\begin{equation}
 c_m
 =
 -\frac{\sin(m\theta_F)}{\pi m},
 \qquad m\ne0 .
 \label{S:eq:atomiccm}
\end{equation}
Thus, the atomic limit provides an explicit realization of
an $r=1$ Fourier tail.

More generally, suppose that a $2\pi$-periodic hull profile has a
single jump of height $J\ne0$ at $\theta=\theta_0$ and that the rest
part is sufficiently regular. Distributional integration by parts
gives
\begin{equation}
 c_m
 =
 \frac{J}{2\pi i m}e^{-im\theta_0}
 +o(m^{-1}).
 \label{S:eq:singleJump}
\end{equation}
A simple jump therefore provides one concrete mechanism for an
$r=1$ tail. However, we emphasize that neither \textbf{Theorem~S1} nor the general Fourier-tail condition requires the algebraic decay to originate from a jump.

For multiple jumps, the leading Fourier coefficient contains an
interference factor. Let $J_a(E_F)$ denote the signed height of the
$a$-th jump and let $\theta_a(E_F)$ denote its position. For finitely
many jumps,
\begin{equation}
 c_m
 =
 \frac{1}{2\pi i m}
 \sum_{a=1}^{n_a}
 J_a(E_F)e^{-im\theta_a(E_F)}
 +o(m^{-1}).
 \label{S:eq:multipleJump}
\end{equation}
Different jump contributions can interfere destructively, and
selected Fourier coefficients may vanish even though the
overall envelope remains proportional to $1/|m|$. Importantly,
\textbf{Theorem~S1} does not require a lower bound at every large
Fourier index. It is sufficient that the normalized Fourier
amplitude remain bounded away from zero along one subsequence
realizing the arithmetic resonance exponent.

The atomic-limit Aubry-Andr\'e model mentioned above gives a direct demonstration of this point. Let $\{q_j\}$ be a sequence of distinct positive integers
realizing the arithmetic resonance exponent,
\begin{equation}
 -\frac{\log\|q_j\beta\|_{\mathbb Z}}{\log q_j}
 \longrightarrow
 \omega=\mu(\beta)-1 .
 \label{S:eq:qRealizing}
\end{equation}
Equation~\eqref{S:eq:atomiccm} gives
\begin{equation}
 q_j|c_{q_j}|
 =
 \frac{|\sin(q_j\theta_F)|}{\pi}.
 \label{S:eq:atomicNondeg}
\end{equation}
The indices $q_j$ are fixed entirely by the arithmetic of $\beta$,
whereas the interference factor depends on the Fermi energy.

For any fixed sequence of distinct integers $\{q_j\}$, Weyl's metric
equidistribution theorem~\cite{Weyl1916} states that
\begin{equation}
 \frac{q_j\theta_F}{2\pi}
 \pmod 1
\end{equation}
is uniformly distributed for Lebesgue-almost every $\theta_F$.
Consider, for example, any finite union of intervals on which
$|\sin(2\pi x)|>1/2$. This set has a positive measure and, therefore, uniform
distribution implies that it is visited by
$q_j\theta_F/(2\pi)$ infinitely often. Hence, an infinite subsequence
satisfies
\begin{equation}
 |\sin(q_j\theta_F)|>\frac12 ,
\end{equation}
and consequently
\begin{equation}
 |c_{q_j}|
 >
 \frac{1}{2\pi q_j}.
\end{equation}
Thus, the $r=1$ nondegeneracy condition holds for
Lebesgue-almost every $E_F$ in the interior of the spectrum.
The exceptional set has zero Lebesgue measure and hence contains no
interval of nonzero length.

Individual Fourier coefficients may still vanish at special
fillings. Such isolated zeros do not violate \textbf{Theorem~S1},
because the theorem requires nonvanishing $1/|m|$ weight only along
one infinite arithmetic exponent-realizing subsequence.

\subsection{Genericity for finitely many moving singularities}

The preceding argument is not intrinsically restricted to
$r=1$ or to jump discontinuities. Consider more generally the case that a hull
profile whose leading large-$m$ Fourier asymptotics, for $m>0$, is
generated by finitely many moving singularities,
\begin{equation}
 c_m(E_F)
 =
 m^{-r}
 \sum_{a=1}^{n_a}
 A_a(E_F)e^{-im\theta_a(E_F)}+o(m^{-r}).
 \label{S:eq:generalMultiSingularity}
\end{equation}
Here, the exponent $r$ is fixed by the type of the leading
singularity. A simple jump gives $r=1$, while continuous cusps or
singularities in derivatives can produce different exponents. For $E_F$ under consideration, we take
$\theta_a(E_F)\pmod{2\pi}$ to be the distinct hull positions of the singularities
contributing at order $m^{-r}$ and $A_a(E_F)\neq0$ to be the
corresponding leading coefficients. Thus, the leading term in
Eq.~\eqref{S:eq:generalMultiSingularity} is not identically
zero.

Although the leading Fourier envelope of $c_m(E_F)$ scales as $|m|^{-r}$, the relative singularity positions can produce
destructive interference that suppresses or cancels selected
coefficients at particular $m$ and $E_F$. Define
\begin{equation}
 \phi_a(E_F)
 :=
 \frac{
 \theta_{a+1}(E_F)-\theta_1(E_F)
 }{2\pi},
 \qquad
 a=1,\ldots,n_a-1 ,
 \label{S:eq:relativePhases}
\end{equation}
and
\begin{equation}
 \boldsymbol{\phi}(E_F)
 :=
 (\phi_1,\ldots,\phi_{n_a-1}) .
\end{equation}
After removing the irrelevant overall phase
$e^{-im\theta_1(E_F)}$, the leading normalized Fourier amplitude is
the trigonometric polynomial
\begin{equation}
 \mathcal{P}_{E_F}(\bm x)
=
 A_1(E_F)
 +
 \sum_{a=2}^{n_a}
 A_a(E_F)e^{-2\pi i x_{a-1}}
 \label{S:eq:torusPolynomial}
\end{equation}
on the torus $\mathbb T^{n_a-1}$. Note that $\mathcal{P}_{E_F}(m\boldsymbol{\phi})=m^r e^{im\theta_1(E_F)}c_m(E_F)+o(1)$.

Exact symmetries or structural constraints may impose homogeneous integer relations among the relative phases, namely, $\mathbf h\cdot\boldsymbol{\phi}=0\pmod{1}$ for some $
\mathbf h\in\mathbb Z^{n_a-1}\setminus\{{\mathbf 0}\}$ that holds throughout the Fermi-energy interval under consideration. In this case, the physically accessible interference phases are restricted to a lower-dimensional subtorus of $\mathbb T^{n_a-1}$. For example, if symmetry pairs two relative phases as
\begin{equation}
\phi_a(E_F)=\frac12+\delta_a(E_F),
\qquad
\phi_{a'}(E_F)=\frac12-\delta_a(E_F),
\end{equation}
then $\phi_a+\phi_{a'}=1=0\pmod{1}$. Hence, the pair contains only one independent phase degree of freedom since $\phi_{a'}=-\phi_a\pmod{1}$ [see Figs. 1(c,d) in the main text, for instance].

We incorporate such exact relations before imposing the genericity condition below. Let $\mathcal T_s\subset\mathbb T^{n_a-1}$ denote the resulting $d_s$-dimensional symmetry-allowed subtorus. We introduce $d_s$ independent phase coordinates
$\widetilde{\boldsymbol{\phi}}(E_F)\in\mathbb T^{d_s}$,
each defined modulo one. Let
$B_s$ be an $(n_a-1)\times d_s$ integer matrix whose columns specify how the $d_s$ independent phases enter the original relative phases. Then,
\begin{equation}
\boldsymbol{\phi}(E_F)
=
B_s\widetilde{\boldsymbol{\phi}}(E_F)
\pmod{\mathbb Z^{n_a-1}}.
\label{S:eq:effectivePhase}
\end{equation}
Restricting $\mathcal P_{E_F}$ to this symmetry-allowed subtorus defines
\begin{equation}
\widetilde{\mathcal P}_{E_F}(\bm x)
:=
\mathcal P_{E_F}(B_s\bm x),
\qquad
\bm x\in\mathbb T^{d_s}.
\label{S:eq:effectivePolynomial}
\end{equation}
Consequently,
\begin{equation}
 m^r e^{im\theta_1(E_F)}c_m(E_F)
 =
 \widetilde{\mathcal P}_{E_F}
 \!\left(m\widetilde{\boldsymbol{\phi}}(E_F)\right)
 +o(1).
 \label{S:eq:effectiveAsymptotics}
\end{equation}

\begin{figure}[h]
    \centering
    \includegraphics[width=0.7\textwidth]{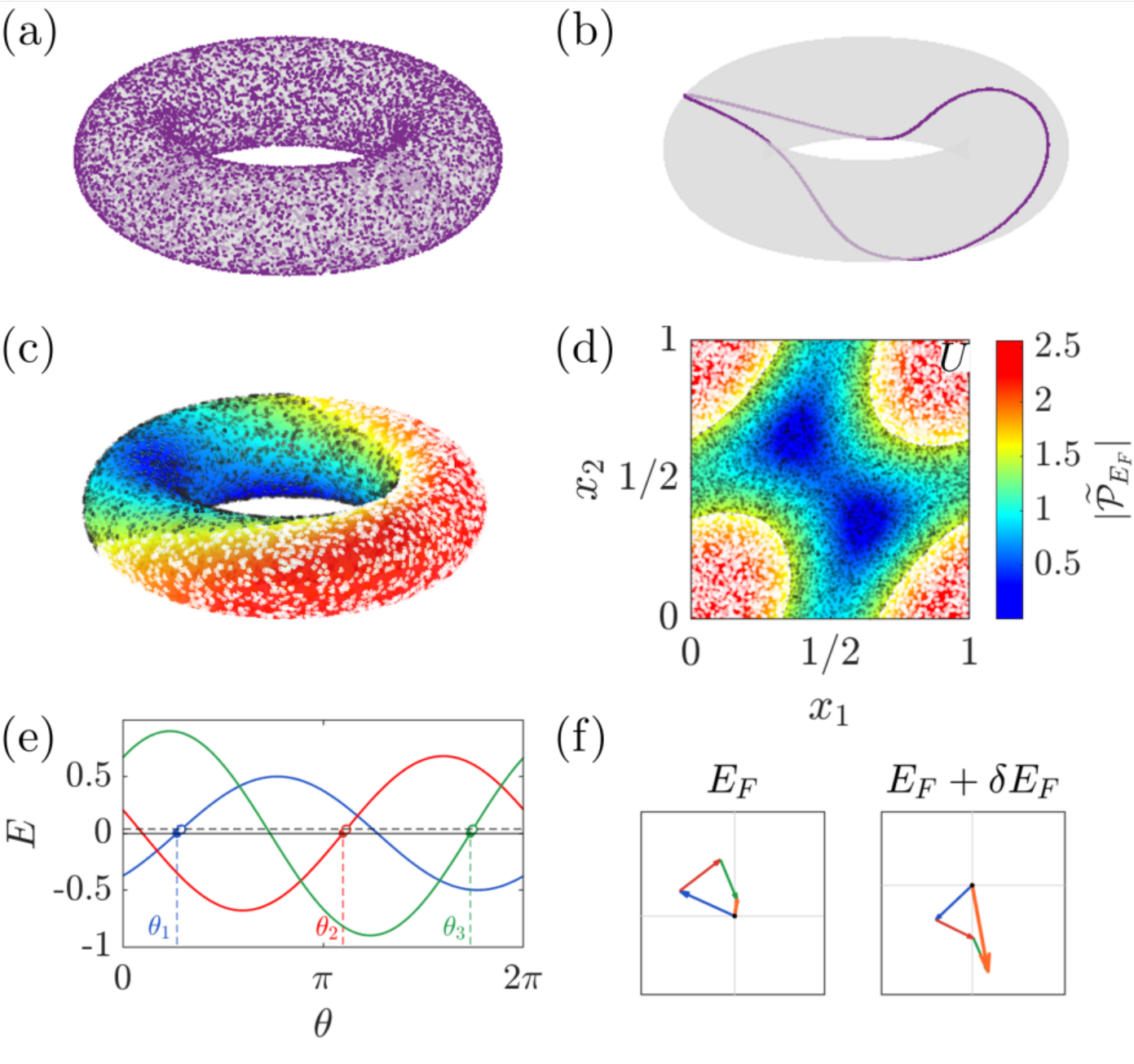}
    \caption{Schematic illustration of the genericity argument for the nondegeneracy condition. (a,b) Geometric consequence of Eq.~\eqref{S:eq:noPhaseLocking} on the effective phase torus after exact symmetry constraints have been incorporated. In the absence of additional integer locking among the independent phase coordinates, the arithmetic orbit explores the full two-dimensional effective torus (a), whereas an additional locking condition, illustrated by $2\widetilde{\phi}_1-\widetilde{\phi}_2=0\pmod 1$, confines the orbit to a one-dimensional subtorus (b), thereby permitting persistent suppression of the resonant Fourier weight. (c,d) Effective torus of two independent coordinates $x_1$ and $x_2$. The surface color represents the magnitude of the effective trigonometric polynomial $|\widetilde{\mathcal{P}}_{E_F}(\boldsymbol{x})|$, while the points show the arithmetic orbit $q_j\widetilde{\boldsymbol{\phi}}(E_F)$ modulo $\mathbb{Z}^2$. White points indicate resonances satisfying $|\widetilde{\mathcal{P}}_{E_F}|\ge 2A_*$, i.e., those lying inside the open set $U$, whereas black points lie outside $U$. (e) Localized branches $E_a(\theta)$ crossing Fermi energy at different $\theta$ values, $\theta_a$ with $a=1,2$ and $3$. Black solid horizontal line represents $E_F$, while dashed line represents shifted Fermi energy, $E_F+\delta E_F$. (f) The corresponding contributions $A_a(E_F)e^{-iq_j\theta_a(E_F)}$ interfere at an exponent-realizing Fibonacci denominator $q_j=2584$ of the golden ratio frequency. The orange arrow represents the resultant interference amplitude. (Left) A strongly cancelled mode at $E_F$. (Right) After $E_F\to E_F+\delta E_F$, the same $q_j$ acquires modified amplitudes and, in particular, amplified phase shifts $q_j\delta\theta_a$, altering the interference.
    }
    \label{fig: S_condition2}
\end{figure}


Having incorporated the exact symmetry constraints into the effective $d_s$-torus, a useful sufficient condition for nondegeneracy [Eq.~\eqref{S:eq:tailLower}] is the absence of any additional integer locking among the independent phase coordinates. Suppose that the components of $\widetilde{\boldsymbol{\phi}}(E_F)$ admit real-analytic lifts on the Fermi-energy interval under consideration and that, for every nonzero integer vector $\bm h\in\mathbb Z^{d_s}$,
\begin{equation}
 \bm h\cdot\widetilde{\boldsymbol{\phi}}(E_F)
 \neq {\rm const}.
 \label{S:eq:noPhaseLocking}
\end{equation}
Eq.~\eqref{S:eq:noPhaseLocking} does not exclude exact symmetry relations already incorporated into the definition of the effective torus, but excludes any unexpected additional structural relation that locks a nontrivial integer combination of the remaining independent phases as the filling is varied. Note that for a fixed singularity type (same $r$) arising from an analytic hull Hamiltonian, both the singularity position $\theta_a(E_F)$ and the corresponding amplitude $A_a(E_F)$ are real and vary locally analytically with $E_F$, as long as the same singularity branch persists and remains
transverse to the Fermi level~\cite{Kato1995}. Isolated exceptional Fermi energies at which this local analytic description fails, such as tangential crossings or branch collisions, may be excluded as a measure-zero set and hence do not affect the almost-everywhere conclusion below.

As a consequence of Eq.~\eqref{S:eq:noPhaseLocking}, $q_j\widetilde{\boldsymbol{\phi}}(E_F)\pmod{\mathbb Z^{d_s}}$ is uniformly distributed on $\mathbb T^{d_s}$ for Lebesgue-almost every $E_F$ under consideration and the fixed exponent-realizing sequence
$\{q_j\}$. In detail, we first fix
$\bm h\ne\bm0$ and define
\begin{equation}
 y_{\bm h}(E_F)
 :=
 \bm h\cdot\widetilde{\boldsymbol{\phi}}(E_F).
\end{equation}
Since $y_{\bm h}$ is non-constant and real analytic, its derivative
can vanish only at isolated points; otherwise, it vanishes everywhere due to the identity theorem, and hence contradicts Eq.~\eqref{S:eq:noPhaseLocking}. 
Away from those points, $E_F\mapsto y_{\bm h}(E_F)$ is locally
invertible. Thus, the preimage of the measure-zero exceptional set remains of measure zero. Hence, Weyl's metric equidistribution theorem applied to the fixed exponent-realizing sequence
$\{q_j\}$ implies
\begin{equation}
 q_j y_{\bm h}(E_F)
 \pmod 1
\end{equation}
to be uniformly distributed for Lebesgue-almost every not only $y_{\bm h}$ but also $E_F$.
Since the set
$\mathbb Z^{d_s}\setminus\{\bm0\}$ is countable, the union of measure-zero exceptional
sets for each $\bm h$ is also measure-zero, and hence can be removed simultaneously for all $\bm h\neq\bm 0$.
The multidimensional Weyl criterion~\cite{Weyl1916,KuipersNiederreiter1974} then gives
\begin{equation}
 q_j\widetilde{\boldsymbol{\phi}}(E_F)
 \pmod{\mathbb Z^{d_s}}
\end{equation}
uniformly distributed on $\mathbb T^{d_s}$ for
Lebesgue-almost every $E_F$ [see Fig.~\ref{fig: S_condition2}(a)].
In detail, the criterion requires
\begin{equation}
 \lim_{N\to\infty}
 \frac{1}{N}
 \sum_{j=1}^{N}
 e^{2\pi i \bm h\cdot(q_j\widetilde{\boldsymbol{\phi}})}
 =0
\end{equation}
for every $\bm h\in\mathbb Z^{d_s}\setminus\{\bm0\}$, which follows directly from the one-dimensional uniform distribution of
$q_jy_{\bm h}(E_F)=q_j\bm h\cdot\widetilde{\boldsymbol{\phi}}(E_F)\pmod 1$
established above. 

Now let us see a consequence of this uniformity. Since $\widetilde{\mathcal{P}}_{E_F}$ is not identically zero for any such $E_F$ under consideration, there exists a constant $A_*(E_F)>0$ and a nonempty open subset
$U(E_F)\subset\mathbb T^{d_s}$ such that
\begin{equation}
 \vert \widetilde{\mathcal{P}}_{E_F}(\bm x)\vert>2A_*(E_F)
\end{equation}
for all $\bm x\in U(E_F)$ [see Figs.~\ref{fig: S_condition2}(c,d)].
Since the set $U(E_F)$ has positive Haar measure, uniform distribution guarantees infinitely many indices $j$ for which
\begin{equation}
 q_j\widetilde{\boldsymbol{\phi}}(E_F)
 \pmod{\mathbb Z^{d_s}}
 \in U(E_F).
\end{equation}
Note that for all sufficiently large $q_j$, $|o(1)|<A_*(E_F)$ in Eq.~\eqref{S:eq:effectiveAsymptotics}. Thus, the triangle inequality yields,
\begin{align}
 q_j^r |c_{q_j}(E_F)|
 &=
 \left|
\widetilde{\mathcal{P}}_{E_F}\!\left(q_j\widetilde{\boldsymbol{\phi}}(E_F)\right)
 + o(1)
 \right|
 \nonumber\\
 &\ge
 \left|
\widetilde{\mathcal{P}}_{E_F}\!\left(q_j\widetilde{\boldsymbol{\phi}}(E_F)\right)
 \right|
 - |o(1)|
 \nonumber\\
 &>
 2A_*(E_F)-A_*(E_F)
 =
 A_*(E_F).
\end{align}
Hence,
\begin{equation}
 |c_{q_j}(E_F)|>A_*(E_F)q_j^{-r}
 \label{S:eq:metricNondegeneracy}
\end{equation}
along an infinite subsequence.

When the original sequence $\{q_j\}$ is chosen to realize
$\omega$, every subsequence of $\{q_j\}$ still satisfies
Eq.~\eqref{S:eq:qRealizing}. Equation~\eqref{S:eq:metricNondegeneracy}
is therefore precisely the second condition required in
\textbf{Theorem~S1}. Hence, after exact symmetry constraints are incorporated into the effective torus, the nondegeneracy condition holds for almost every $E_F$ under the sufficient no-additional-locking condition \eqref{S:eq:noPhaseLocking}.

This formulation also clarifies the exceptional cases. An exact symmetry or structural relation among the original singularity positions does not by itself imply persistent suppression of the resonant Fourier weight; rather, such a relation reduces the dimension of the effective interference torus and should be incorporated into its definition. Persistent suppression requires an additional restriction within the effective torus that prevents the exponent-realizing orbit from visiting a region of nonvanishing effective Fourier amplitude, or an exact cancellation for which the effective trigonometric polynomial vanishes identically. In the latter case, the nominal exponent $r$ does not represent the leading Fourier tail and must be reidentified. In the absence of such additional locking, systematic destructive interference throughout the complete resonance hierarchy does not occur generally.

\subsection{Robustness against finite hopping}

At finite hopping, the atomic-site levels are hybridized into localized eigenstates. Thus, one should consider the localized spectral branches rather than individual onsite levels. Then, a natural source of a jump in the hull charge profile becomes a
localized spectral level crossing the Fermi energy as the hull phase
is varied. Let $E_a(\theta)$ denote an isolated spectral branch. The
position of the corresponding crossing satisfies
\begin{equation}
 E_a[\theta_a(E_F)]= E_F.
 \label{S:eq:crossingCondition}
\end{equation}
For an isolated eigenvalue branch $E_a(\theta)$ with normalized
eigenstate $|\psi_a(\theta)\rangle$, the Hellmann--Feynman theorem
gives
\begin{equation}
 \partial_\theta E_a(\theta)= -\lambda
 \sum_j
 |\langle j|\psi_a(\theta)\rangle|^2
 \sin(2\pi\beta j+\theta).
 \label{S:eq:HFtheta}
\end{equation}
At a Fermi-level crossing, there is generically no symmetry forcing
this derivative to vanish. Note that in the atomic limit, Eq.~\eqref{S:eq:HFtheta} reduces to $\partial_\theta E_a(\theta)\simeq -\lambda\sin(2\pi\beta j_a+\theta)$. Thus, a vanishing derivative requires the crossing to occur near an
extremum of the local quasiperiodic potential in the atomic limit. Since tangential crossings require an additional tuning, a
transverse crossing,
\begin{equation}
 \partial_\theta
 E_a[\theta_a(E_F)]
 \ne0,
\end{equation}
is expected for a generic Fermi energy.

For a transverse crossing, the implicit-function theorem gives
\begin{equation}
 \frac{d\theta_a}{dE_F}
 =
 \frac{1}{
 \partial_\theta E_a[\theta_a(E_F)]
 }.
 \label{S:eq:crossingMotion}
\end{equation}
The arithmetic resonance indices $q_j$ are unchanged when $E_F$ is
varied, whereas the interference phases in
Eq.~\eqref{S:eq:multipleJump} move. A small variation
$\delta E_F$ changes the phase $q_j\theta_a$ associated with the $a$-th jump by
\begin{equation}
 \delta[q_j\theta_a]
 \simeq
 \frac{
 q_j\,\delta E_F
 }{
 \partial_\theta E_a[\theta_a(E_F)]
 }.
 \label{S:eq:phaseSensitivity}
\end{equation}
Since $q_j\to\infty$ along the resonance hierarchy, an arbitrarily
small but fixed nonzero change of $E_F$ eventually generates an
order-one change of the interference phase. Hence, relative interference
phases become increasingly sensitive to the filling at
high resonance order [see Figs.~\ref{fig: S_condition2}(e,f)].

For multiple crossings, the relevant quantities are the relative
phases. From Eq.~\eqref{S:eq:crossingMotion},
\begin{equation}
 \frac{d}{dE_F}
 [\theta_a(E_F)-\theta_b(E_F)]
 =
 \frac{1}{
 \partial_\theta E_a[\theta_a(E_F)]
 }
 -
 \frac{1}{
 \partial_\theta E_b[\theta_b(E_F)]
 }.
 \label{S:eq:relativeCrossingMotion}
\end{equation}
Thus, an exact locking of the relative crossing positions over a
finite interval requires a functional relation among distinct
crossing velocities. Such a relation is not expected generically and
can instead be enforced by a symmetry or other structural
constraints. 
This implies that finite hopping, although it renormalizes the microscopic positions, amplitudes, and velocities of the localized crossings through
hybridization, does not by itself generate the additional locking required to invalidate the above metric argument. As long as the relevant isolated
transverse branches persist, the essential conclusion remains unchanged. Accordingly, the nondegeneracy condition Eq.~\eqref{S:eq:tailLower} is robust against finite hopping.

The level-crossing construction is specific to a convenient
localized realization of the $r=1$ tail without additional symmetry, but the metric argument
itself is neither restricted to localized states nor to simple
jumps. If an extended or critical regime produces a finite set of moving singularities with the asymptotic structure
of Eq.~\eqref{S:eq:generalMultiSingularity}, the same metric
nondegeneracy argument applies.

We do not, however, assume that every
gapless extended, critical, or localized phase should admit such a finite
singularity representation. For this reason, the above finite-singularity analysis is a
sufficient genericity argument rather than an assumption entering the
general theorem. \textbf{Theorem~S1} retains the Fourier-tail bound
and the nondegeneracy condition along an arithmetic
exponent-realizing subsequence as explicit assumptions for arbitrary
$r>1/2$. Stable gapped sectors, whose hull contribution is analytic
rather than algebraically singular, are treated separately in Sec.~\ref{S:sec:stablegap}.

\subsection{Logarithmic weakening of the nondegeneracy condition}
\label{S:subsec:logNondegeneracy}

The fixed-amplitude nondegeneracy condition Eq.~\eqref{S:eq:tailLower} used in \textbf{Theorem~S1} is physically natural but stronger than required for determining the logarithmic scaling exponent since it excludes subpower corrections such as $|c_{q_j}|\sim q_j^{-r}/\log q_j$. We emphasize that the conclusion of \textbf{Theorem~S1} remains valid if the fixed-amplitude nondegeneracy condition is replaced by the weaker requirement that there exists an exponent-realizing sequence $\{q_j\}$ satisfying
\begin{equation}
\lim_{j\to\infty}\frac{-\log |c_{q_j}|}{\log q_j}=r,
\label{S:eq:logNondegeneracy}
\end{equation}
or, equivalently,
\begin{equation}
|c_{q_j}|=q_j^{-r+o(1)}.
\label{S:eq:logNondegeneracyEquivalent}
\end{equation}
Thus, subpower corrections such as $|c_{q_j}|\sim q_j^{-r}/\log q_j$ are allowed. The stronger condition Eq.~\eqref{S:eq:tailLower}, together with the global upper bound $|c_m|\le B|m|^{-r}$, immediately implies Eq.~\eqref{S:eq:logNondegeneracy}.

To see that the conclusion of \textbf{Theorem~S1} is unchanged under Eq.~\eqref{S:eq:logNondegeneracy}, define
\begin{equation}
b_j:=-\frac{\log |c_{q_j}|}{\log q_j},
\qquad b_j\to r.
\label{S:eq:bjLogNondegeneracy}
\end{equation}
The lower bound on the logarithmic exponent is unchanged since that part of the proof uses only the global Fourier bound. For the matching reverse inequality, retain the definitions
\[
a_j:=-\frac{\log\|q_j\beta\|_{\mathbb Z}}{\log q_j}\to\omega,
\qquad
K_j:=2\|q_j\beta\|_{\mathbb Z}.
\]
The single $q_j$-th harmonic then gives
\begin{equation}
Z(K_j)\ge |c_{q_j}|^2=q_j^{-2b_j}.
\label{S:eq:logSingleMode}
\end{equation}
Since $\log K_j=\log 2-a_j\log q_j<0$ for sufficiently large $j$, Eq.~\eqref{S:eq:logSingleMode} yields
\begin{equation}
\frac{\log Z(K_j)}{\log K_j}
\le
\frac{-2b_j\log q_j}{\log 2-a_j\log q_j}
=
\frac{2b_j}{a_j-\dfrac{\log 2}{\log q_j}}
\longrightarrow
\frac{2r}{\omega}.
\label{S:eq:logUpperBound}
\end{equation}
The remaining subsequence argument is identical to that in the proof of \textbf{Theorem~S1}, and combining this result with the unchanged opposite bound gives
\[
\liminf_{K\to0^+}\frac{\log Z(K)}{\log K}
=
\frac{2r}{\omega}.
\]
Hence, Eq.~\eqref{S:eq:logNondegeneracy} is sufficient for \textbf{Theorem~S1}. Nevertheless, the stronger fixed-amplitude condition [Eq.~\eqref{S:eq:tailLower}] is natural for the finite-singularity setting considered above, where the metric equidistribution argument gives $|c_{q_j}|>A_*q_j^{-r}$ along an infinite exponent-realizing subsequence.

\section{Stable-gap analyticity and Fermi-level dominance}
\label{S:sec:stablegap}

\subsection{Proof of stable-gap analyticity and Fermi-level dominance}
Here, we show that a stable separation of the deep occupied sector makes its long-wavelength contribution asymptotically negligible, leaving the scaling controlled by the states near the Fermi level.
Let $E_F$ be the Fermi energy and choose $E_g<E_F$ inside a spectral gap that remains open for every real hull phase, which we refer to as a stable gap [see Fig.~\ref{fig: S_stable}(a)]. Denote by $P_{E\le E_F}$ the Fermi projector, by $P_{E<E_g}$ the projector onto the deep occupied spectral sector below $E_g$, and by $P_{E_g\le E\le E_F}$ the complementary occupied shell. Then
\begin{equation}
 P_{E\le E_F}=P_{E_g\le E\le E_F}+P_{E<E_g}.
 \label{S:eq:projectorSplit}
\end{equation}
Their local diagonal matrix elements satisfy $n_j=n_j^{\rm shell}+n_j^{\rm deep}$. We define the deep contribution by
\begin{equation}
 n_j^{\rm deep}:=\langle j|P_{E<E_g}\vert j\rangle.
 \label{S:eq:densitySplit}
\end{equation}
Translation covariance gives a deep-sector hull function $F_{\rm deep}$ through
\begin{equation}
 n_j^{\rm deep}(\theta)=F_{\rm deep}(\theta+2\pi\beta j).
 \label{S:eq:FdeepDef}
\end{equation}

Assume that $H(\theta)$ admits an analytic continuation to complex hull phase in the strip $|\operatorname{Im}\theta|<\rho_0$ and that the deep spectral sector remains separated by a gap from the rest of the spectrum throughout a smaller closed strip $|\operatorname{Im}\theta|\le\rho<\rho_0$. Let $\Gamma$ be a positively oriented closed contour in the complex energy $z$ plane with the following uniform property: for every $\theta$ in that smaller strip, $\Gamma$ lies in the resolvent set of $H(\theta)$, encloses the continuation of the spectral sector below $E_g$, and encloses no spectrum belonging to the complementary sector [see Fig.~\ref{fig: S_stable}(b)]. The Riesz projector is then
\begin{equation}
 P_{E<E_g}(\theta)=\frac{1}{2\pi i}\oint_\Gamma
 dz\,[z-H(\theta)]^{-1}.
 \label{S:eq:Riesz}
\end{equation}
The existence of one such contour for the entire strip is the precise stability input used here. Since the resolvent $[z-H(\theta)]^{-1}$ is analytic in $\theta$ and uniformly bounded on the compact contour-strip domain, Eq.~\eqref{S:eq:Riesz} implies that $P_{E<E_g}(\theta)$ and $F_{\rm deep}(\theta)$ are analytic and bounded in the strip.
\begin{figure}[h]
    \centering
    \includegraphics[width=0.7\textwidth]{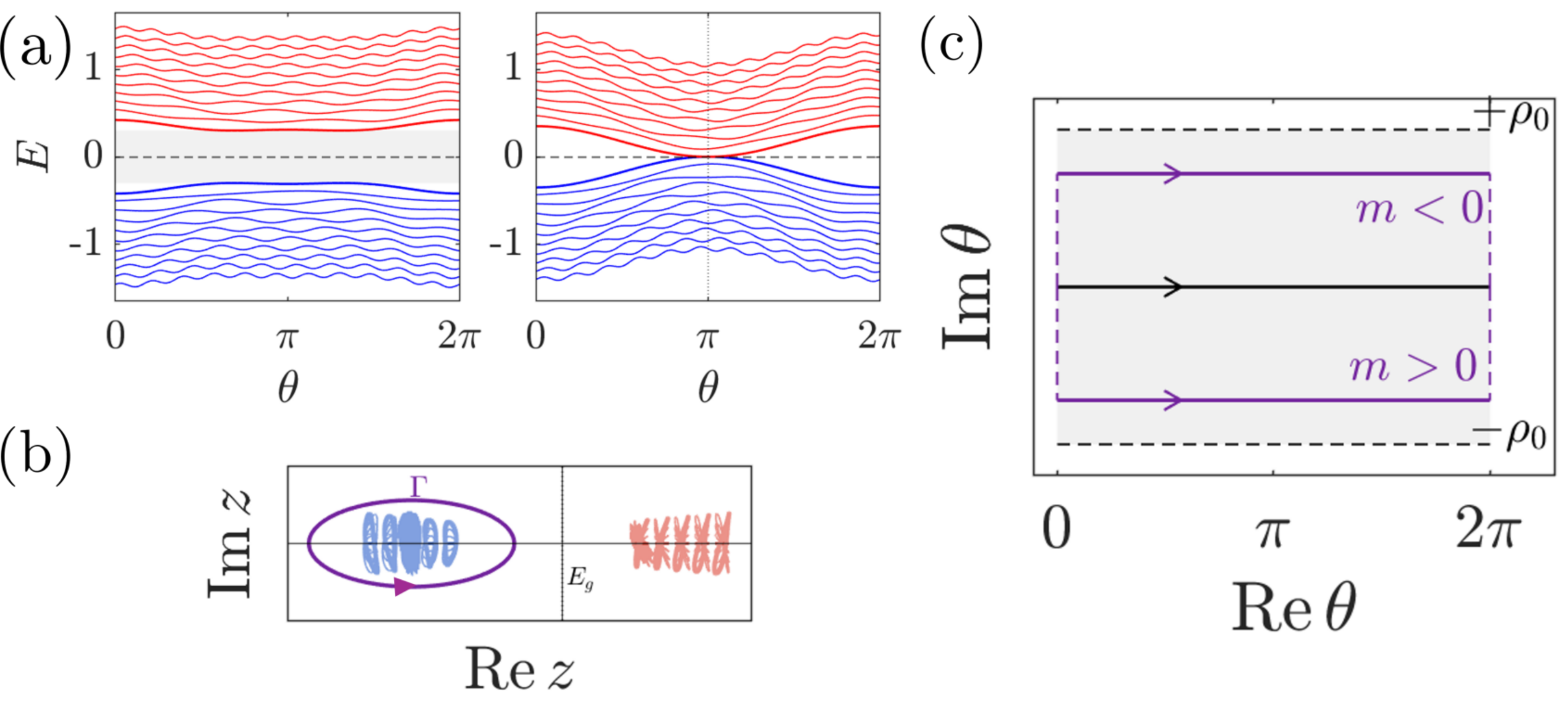}
    \caption{Schematic illustration of the stable-gap analyticity argument. (a) A hull-stable spectral gap remains open for all real $\theta$ (left), in contrast to a phase-specific gap that closes as $\theta$ is varied (right). (b) For a stable gap, a single positively-oriented contour $\Gamma$ encloses the deep occupied spectral sector (blue) and excludes the complementary sector (red) throughout the strip after analytic continuation to $|\operatorname{Im}\theta|\leq\rho$. This gives the analytic Riesz projector in Eq.~\eqref{S:eq:Riesz}. (c) The Fourier contour (black) associated with Eq.~\eqref{S:eq:deepFourier}. The contour shifted to lower half plane for $m>0$ and upper half plane for $m<0$, leading to the exponential Fourier bound in Eq.~\eqref{S:eq:expDecay}.
    }
    \label{fig: S_stable}
\end{figure}

Define the Fourier coefficients $d_m$ of the deep hull function by
\begin{equation}
 F_{\rm deep}(\theta)=\sum_{m\in\Z}d_m e^{im\theta}.
 \label{S:eq:deepFourier}
\end{equation}
Shifting the Fourier integration contour vertically by the strip width gives constants $C>0$ and $\rho>0$, independent of $m$, such that
\begin{equation}
 |d_m|\le C e^{-\rho|m|}.
 \label{S:eq:expDecay}
\end{equation}
For finite $\mu$ and any fixed $\epsilon>0$, Eq.~\eqref{S:eq:eventualLower} implies that every sufficiently small-$K$ resonance satisfies
\begin{equation}
 |m|\gtrsim K^{-1/(\mu-1+\epsilon)}.
 \label{S:eq:deepIndexBound}
\end{equation}
Here, $A(K)\gtrsim B(K)$ means $A(K)\ge C_0B(K)$ for some constant $C_0>0$ independent of sufficiently small $K$. Therefore, the integrated deep-sector weight
\begin{equation}
 Z_{\rm deep}(K):=
 \sum_{m\ne0:\,\normZ{m\beta}<K}|d_m|^2
 \label{S:eq:ZdeepDef}
\end{equation}
obeys
\begin{equation}
 Z_{\rm deep}(K)\le C_\epsilon
 \exp\left[-a_\epsilon K^{-1/(\mu-1+\epsilon)}\right]
 \label{S:eq:deepSuper}
\end{equation}
for positive constants $C_\epsilon$ and $a_\epsilon$. Hence, $Z_{\rm deep}(K)=o(K^p)$ as $K\to0^+$ for every $p>0$.

Let $s_m$ denote the Fourier coefficients of the shell contribution, so the total coefficient is $c_m=s_m+d_m$, and define
\begin{equation}
 Z_{\rm shell}(K):=
 \sum_{m\ne0:\,\normZ{m\beta}<K}|s_m|^2.
\end{equation}
The shell projector is bounded, so its local hull profile is bounded on the real hull and therefore square integrable. Parseval's identity gives
\begin{equation}
 \sum_{m\in\Z}|s_m|^2<\infty,
 \label{S:eq:shellL2}
\end{equation}
and in particular $Z_{\rm shell}(K)$ is uniformly bounded as $K\to0^+$. The cross term is then bounded by the Cauchy--Schwarz inequality,
\begin{equation}
 \left|\sum_{\normZ{m\beta}<K}s_m d_m^*\right|
 \le\sqrt{Z_{\rm shell}(K)Z_{\rm deep}(K)},
 \label{S:eq:crossBound}
\end{equation}
where $d_m^*$ denotes complex conjugation. Since $Z_{\rm deep}(K)$ is superalgebraically small, the cross term is also superalgebraically small. Hence any nontrivial power-law exponent of the full charge distribution is equal to the shell exponent. This provides an analytical framework for understanding the recent findings in the Aubry--Andr\'e and generalized Aubry--Andr\'e models~\cite{SakaiAritaOhtsuki2022,XiangJeonSakai2026}.

\subsection{Application to generalized Aubry-Andr\'e model}
As an explicit application of the analytic-hull argument, let us consider
the generalized Aubry-Andr\'e model~\cite{GaneshanPixleyDasSarma2015}
\begin{align}
 H(\theta)
 &=
 -t\sum_{j\in\Z}
 \left(
 f_j^\dagger f_{j+1}
 +f_{j+1}^\dagger f_j
 \right)
 +\sum_{j\in\Z}
 V(\theta+2\pi\beta j)\,
 f_j^\dagger f_j,
 \label{S:eq:GAAH}\\
 V(\vartheta)
 &=
 \frac{\lambda\cos\vartheta}
 {1-\eta\cos\vartheta},
 \qquad |\eta|<1,
 \label{S:eq:GAApotential}
\end{align}
where $t$ is the nearest-neighbor hopping amplitude, $\lambda$ is the onsite potential strength, and $\eta$ is the deformation
parameter. The Aubry-Andr\'e model is recovered at
$\eta=0$.

We now verify the complex-hull analyticity assumption for the generalized Aubry--Andr\'e model by showing that the Hamiltonian in Eq.~\eqref{S:eq:GAAH} is analytic in a strip of complex hull phase. It is sufficient to establish the analyticity of the potential in Eq.~\eqref{S:eq:GAApotential}. A complex singularity of $V(\vartheta)$ requires
\begin{equation}
 1-\eta\cos(\vartheta+\mathrm{i} y)=0.
 \label{S:eq:GAAsingularity}
\end{equation}
For real $\eta$ with $|\eta|<1$, the singularities closest to the
real hull occur at an imaginary distance
\begin{equation}
 \rho_\eta
 =
 \operatorname{arcosh}\frac{1}{|\eta|}.
 \label{S:eq:GAAstrip}
\end{equation}
Thus, the onsite hull function is analytic in the strip
\begin{equation}
 |\operatorname{Im}\vartheta|<\rho_\eta.
\end{equation}
Consequently, all stable-gapped phases of the generalized Aubry--Andr\'e model exhibit class-I hyperuniform charge distribution, with superalgebraically suppressed long-wavelength fluctuations, whereas in gapless phases the asymptotic scaling is governed by the states near the Fermi level.

For this standard example of quasiperiodic system, the exponential decay of the Fourier coefficients can also be seen directly. For $\eta\ne0$,
define
\begin{equation}
 r_\eta
 :=
 \frac{1-\sqrt{1-\eta^2}}{\eta},
 \qquad |r_\eta|<1.
 \label{S:eq:GAAr}
\end{equation}
Since
\begin{equation}
 \eta=\frac{2r_\eta}{1+r_\eta^2},
\end{equation}
one obtains
\begin{equation}
 \frac{1}{1-\eta\cos\vartheta}
 =
 \frac{1}{\sqrt{1-\eta^2}}
 \sum_{m\in\Z}
 r_\eta^{|m|}e^{im\vartheta}.
 \label{S:eq:GAAFourierDenominator}
\end{equation}
Using
\begin{equation}
 V(\vartheta)
 =
 \frac{\lambda}{\eta}
 \left[
 \frac{1}{1-\eta\cos\vartheta}-1
 \right],
\end{equation}
the nonzero Fourier coefficients $v_m$ of the onsite potential are
\begin{equation}
 v_m
 =
 \frac{\lambda}
 {\eta\sqrt{1-\eta^2}}\,
 r_\eta^{|m|},
 \qquad m\ne0.
 \label{S:eq:GAAFourierCoeff}
\end{equation}
Hence there exist constants $C,\rho>0$ such that
\begin{equation}
 |v_m|\le C e^{-\rho|m|}.
 \label{S:eq:GAAExpDecay}
\end{equation}

For an irrational frequency with finite
$\omega(\beta)=\mu(\beta)-1$ and any fixed $\epsilon>0$, Eq.~\eqref{S:eq:eventualLower} implies that every sufficiently small-$K$ resonance requires
\begin{equation}
 |m|\gtrsim
 K^{-1/(\omega+\epsilon)}.
\end{equation}
Combining this arithmetic constraint with
Eq.~\eqref{S:eq:GAAExpDecay} gives
\begin{equation}
 Z_V(K)
 \lesssim
 \exp\left[
 -c_\epsilon K^{-1/(\omega+\epsilon)}
 \right],
 \label{S:eq:GAAZ}
\end{equation}
where $Z_V(K)$ is the integrated infrared Bragg weight of the onsite
modulation. Thus, $Z_V(K)$ vanishes faster than any power of $K$, and
the generalized Aubry-Andr\'e potential is class-I
hyperuniform.

This example makes the separation between hull regularity and arithmetic explicit. Analyticity of the onsite potential modulation suppresses large-index hull harmonics exponentially, whereas the arithmetic of $\beta$ determines how large a harmonic index is required to reach a given infrared momentum. The above conclusion concerns the onsite potential itself. The hyperuniformity class of the charge density is instead determined by the hull regularity of the corresponding spectral projector and, in a gapless system, by the states near the Fermi level.

\subsection{Why stability of the gap is essential: The case study of particle-hole symmetric quasiperiodic model}

The preceding argument requires a contour $\Gamma$ that remains in the resolvent set throughout the hull and the complex strip. A gap that exists only at one selected real phase does not satisfy this assumption: If a translation of the hull may close the gap, the spectral projector need not possess one uniformly analytic Riesz representation. Thus, a gap present at a single hull phase alone is insufficient to guarantee class-I charge hyperuniformity.

To illustrate why a spectral gap at a single hull phase is insufficient for the stable-gap argument and hence allowing different hyperuniformity-class gapped phase, we introduce a particle-hole-symmetric variation of the Aubry--Andr\'e model,
\begin{equation}
 H=
 -t\sum_j
 \left(
f_j^\dagger f_{j+1}+{\rm H.c.}
 \right)
 +
 \sum_j V_j^{\rm PHS}(\lambda)f_j^\dagger f_j ,
 \label{S:eq:PHSAA}
\end{equation}
with
\begin{equation}
 V_j^{\rm PHS}(\lambda)
 =
 \lambda\cos(2\pi\beta j+\phi_{\rm PHS})
 +
 M_{\rm PHS}(-1)^j ,
 \qquad
 \phi_{\rm PHS}
 =
 \frac{\pi}{2}(1-2\beta),
 \label{S:eq:PHSAApotential}
\end{equation}
Here, $M_{\rm PHS}$ is the mass term. Note that the conventional self-duality of the Aubry-Andr\'e model is absent in the Hamiltonian, Eq.~\eqref{S:eq:PHSAA}. The phase choice in Eq.~\eqref{S:eq:PHSAApotential} gives
\begin{equation}
 V_{1-j}^{\rm PHS}(\lambda)=-V_j^{\rm PHS}(\lambda).
 \label{S:eq:PHSreflection}
\end{equation}
Together with the staggered transformation, this defines the
particle-hole transformation,
\begin{equation}
 \mathcal C f_j\mathcal C^{-1}
 =
 (-1)^j f_{1-j}^\dagger ,
\end{equation}
under which
\begin{equation}
 \mathcal C H\mathcal C^{-1}=-H .
 \label{S:eq:PHSsymmetry}
\end{equation}
Consequently, the single-particle spectrum is symmetric about zero
energy. At half filling, the Fermi energy is pinned to $E_F=0$, and the central gap is
\begin{equation}
 \Delta_0
 =
 E_{L/2+1}-E_{L/2}
 =
 2E_{L/2+1},
 \label{S:eq:centralGap}
\end{equation}
where $L$ is the system size.

Figure~\ref{fig: S1}(a) exhibits $E_{L/2+1}$ and $E_{L/2}$ as functions of $\lambda$. The color represents the fractal dimension of eigenstates, $D_f(L)=-\log \mathcal{I}/\log L$, where $\mathcal{I}$ is inverse participation ratio of eigenstates. As $L$ increases, the fractal dimension approaches unity for extended states, while zero for localized states. We now focus on the large-$\lambda$ regime, where the single-particle states are strongly localized and the spectrum approaches the atomic limit.  In the atomic limit, the eigenenergies are given directly by the onsite potentials $V_j^{\rm PHS}$. Particle-hole symmetry
then implies that the gap around zero energy is
\begin{equation}
 \Delta_0^{\rm at}(\lambda) = 2 \min_j |V_j^{\rm PHS}(\lambda)|.
 \label{S:eq:atomicGap}
\end{equation}
Here, the superscript `$\rm at$' stands for the atomic limit. For finite hopping $t$ and sufficiently large $\lambda$, the low-energy
eigenstates remain strongly localized, and hence the central gap closely follows the same quantity,
\begin{equation}
 \Delta_0(\lambda) \simeq 2 \min_j |V_j^{\rm PHS}(\lambda)|.
 \label{S:eq:gapMinimum}
\end{equation}
This gives rise to the repeated closing and reopening of the central gap. In detail, for each site $j$,
\begin{equation}
 |V_j^{\rm PHS}(\lambda)|= \left| \lambda\cos(2\pi\beta j+\phi_{\rm PHS}) + M_{\rm PHS}(-1)^j
 \right|
\end{equation}
is a piecewise-linear, V-shaped function of $\lambda$ [see Fig.~\ref{fig: S1}(b)]. A given local potential approaches zero near
\begin{equation}
 \lambda_j=-\frac{M_{\rm PHS}(-1)^j}{\cos(2\pi\beta j+\phi_{\rm PHS})},
\end{equation}
whenever $\lambda_j>0$. Since the minimum is taken over
all sites, different localized states successively become the state closest to zero energy as $\lambda$ is varied. The lower envelope of these absolute-value functions consequently develops repeated cusp-like minima, which are reflected in the oscillatory closing and reopening of the half-filling gap [see Fig.~\ref{fig: S1}(b)].

Notably, the resulting central gap at $E_F=0$ is not a stable gap on the quasiperiodic hull since the symmetry is pinned to $\phi_{\rm PHS}$. Specifically, the hull of the system is $S^1\times Z_2$. Without loss of generality, we focus on the trivial sector of $Z_2$ originating from the mass term. Let us introduce a hull displacement $\vartheta$,
\begin{equation}
 V_j^{\rm PHS}(\lambda,\vartheta)=\lambda\cos(2\pi\beta j+\phi_{\rm PHS}+\vartheta) + M_{\rm PHS}(-1)^j.
\end{equation}
Then, the particle-hole transformation gives
\begin{equation}
 \mathcal C H(\vartheta)\mathcal C^{-1} =-H(-\vartheta).
 \label{S:eq:PHShull}
\end{equation}
Thus, exact particle-hole symmetry of the same Hamiltonian is special to the selected phase $\vartheta=0$. Consequently, the central gap would be finite at $\vartheta=0$ while closing at other $\vartheta$,
\begin{equation}
 \Delta_0(0)>0,
 \qquad
 \inf_{\vartheta}\Delta_0(\vartheta)=0.
 \label{S:eq:unstableGap}
\end{equation}
Hence, the gap at $E_F=0$ is phase specific rather than hull stable. This implies that no common Riesz contour can remain separated from the spectrum throughout the hull, and thus the stable-gap analyticity argument is not applicable.

To examine hyperuniformity characteristics of half-filled charge density of this particle-hole symmetric model, we further consider
\begin{equation}
    \gamma_B(R):= \frac{d\ln D_B(R)}{d\ln R},
\end{equation}
where $D_B(R)$ is given by Eq.~\eqref{S:eq:DB}. For a hyperuniform charge density, class I gives
$\gamma_B(R)\to0^-$, class II is marginal with
$\gamma_B(R)\to0$, whereas class III gives
$\gamma_B(R)\to1-\alpha>0$. Thus, a finite positive large-$R$
asymptote of $\gamma_B(R)$ identifies class-III hyperuniformity,
while a persistent negative large-$R$ trend excludes class III, i.e., either class I or II. Notably, we find that the sign of large-$R$ asymptote of $\gamma_B(R)$, say $\gamma_B$, drastically changes as the half-filling gap is closed [see Fig.~\ref{fig: S1}(b)]. This indicates that the hyperuniformity class changes as gap is closed. However, since the closing of the half-filling gap occurs in an instant, this implies that there are class-III gapped phases. Specifically, positive $\gamma_B$ in Fig.~\ref{fig: S1}(b) indicates the class-III hyperuniform charge density of gapped phase [see Fig.~\ref{fig: S1}(c)].
\begin{figure}[h]
    \centering
    \includegraphics[width=0.7\textwidth]{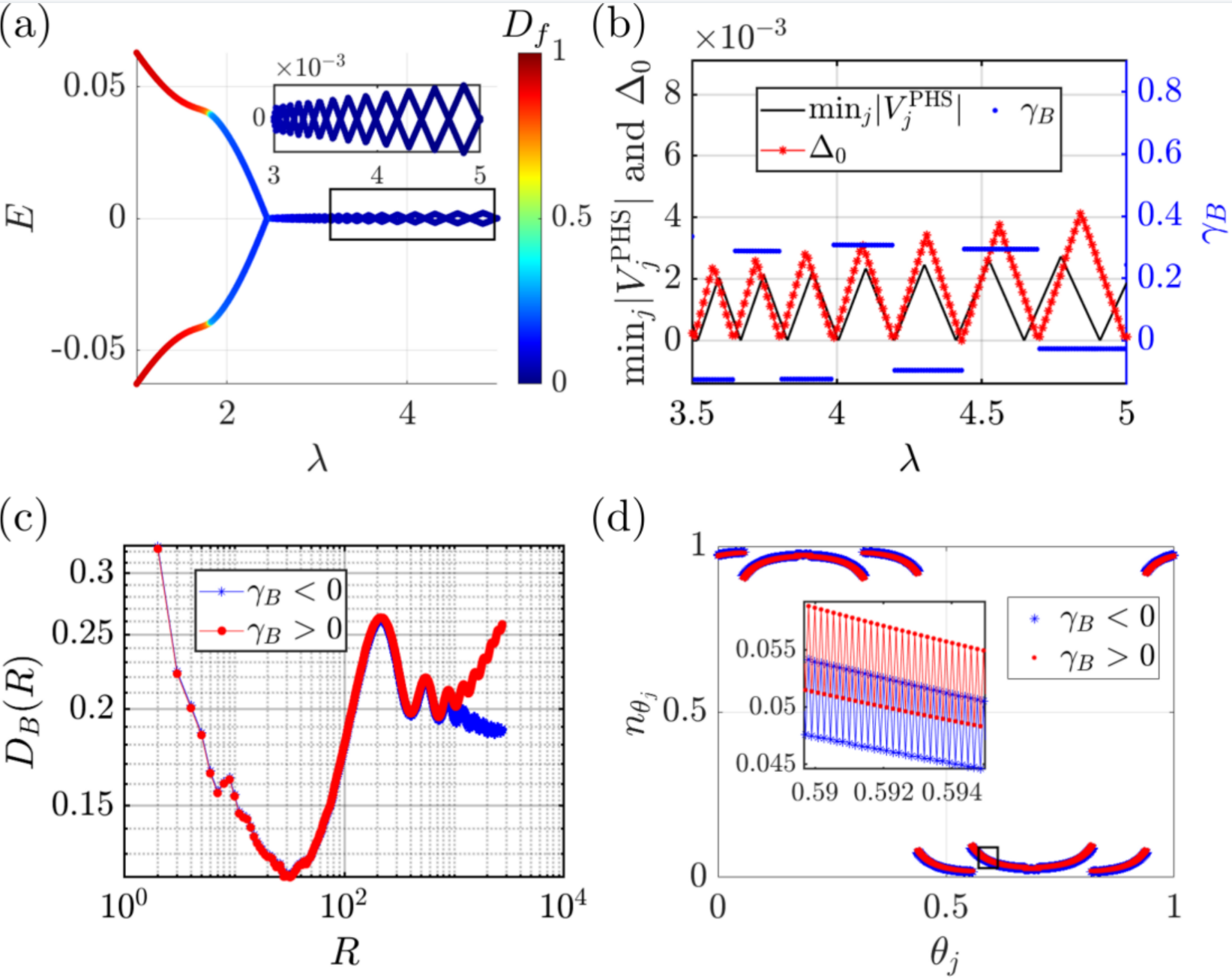}
    \caption{(a) Single-particle eigenenergies, $E_{L/2+1}$ and $E_{L/2}$, around zero energy as a function of $\lambda$. The colors represent the fractal dimension, $D_f$, which is 1 (0) for extended (localized) state in the limit of large $L$. The inset is drawn to emphasize the closing and reopening of half-filling gap at $E_F=0$. (b) The left axis represents the relationship between spectral gap $\Delta_0$ (red) and the minimum value of postentials (black) as functions of $\lambda$. The right axis represents $\gamma_B$ (blue). The sign of $\gamma_B$ drastically changes as half-filling gap is closed. (c) The log-log plot of $D_B(R)$ as a function of window size $R$. The red (blue) represents the case of $\lambda=4.3 \ (4.6)$ with increasing (decreasing) $D_B(R)$, respectively. (d) Hull charge profile $n_{\theta_j}=F(\theta_j)$, where $\theta_j=j\beta \ \mathrm{mod} \ 1$ for both cases of positive and negative $\gamma_B$. The inset is 
    an enlarged view of the region denoted by a black square.
    $\beta=6765/10946$, which is a continued fraction approximant of the golden ratio. The system size is 10946 and $t=1$ is the unit of energy.
    }
    \label{fig: S1}
\end{figure}

Figure~\ref{fig: S1}(d) highlights the $S^1\times Z_2$ hull structure of the particle-hole-symmetric variation of the Aubry--Andr\'e model. The staggered mass term $M_{\rm PHS}(-1)^j$ introduces two $Z_2$ sectors. A one-site translation maps one $Z_2$ sector onto the other while
advancing the continuous hull phase, whereas a two-site translation returns to the same $Z_2$ sector with a further phase shift. Each $Z_2$ sector defines a distinct charge hull profile, and the finite
mismatch between the two sector-resolved profiles appears as jumps when
projected onto the $S^1$ hull coordinate [see the inset of Fig.~\ref{fig: S1}(d)].

\section{Multi-frequency torus-hull generalization}
\subsection{Generalized exceptional-resonance theorem}
\textbf{Theorem~S1} requires only translation covariance, a hull Fourier expansion, and control of the large-index Fourier tail. Thus it applies broadly to quasiperiodic Hamiltonians whenever the relevant local observable satisfies the stated Fourier-tail condition. In this section, we extend \textbf{Theorem~S1} and stable gap result to multi-frequency torus hull.

Let $\Torus:=\mathbb R/(2\pi\Z)$ be the circle of hull phases and consider a $d$-frequency hull $\bm\theta=(\theta_1,\ldots,\theta_d)\in\Torus^d$. Let $\bm\beta=(\beta_1,\ldots,\beta_d)\in\mathbb R^d$ be a frequency vector such that $1,\beta_1,\ldots,\beta_d$ are rationally independent. On the same one-dimensional physical lattice, translation covariance takes the form
\begin{equation}
 T H(\bm\theta)T^{-1}=H(\bm\theta-2\pi\bm\beta),
 \label{S:eq:multiCov}
\end{equation}
where addition of hull phases is understood modulo $2\pi$ in each component.

A covariant local observable can be written as
\begin{equation}
 n_j=F(\bm\theta+2\pi j\bm\beta).
 \label{S:eq:multiHull}
\end{equation}
Let $\bm m=(m_1,\ldots,m_d)\in\Z^d$ denote the integer Fourier index and let $\bm0$ be the zero vector. Define
\begin{equation}
 \overline F:=\frac{1}{(2\pi)^d}\int_{[0,2\pi]^d}
 F(\bm\vartheta)\,d^d\bm\vartheta
\end{equation}
and expand
\begin{equation}
 F(\bm\theta)-\overline F
 =\sum_{\bm m\in\Z^d\setminus\{\bm0\}}
 c_{\bm m}e^{i\bm m\cdot\bm\theta},
 \label{S:eq:multiF}
\end{equation}
where $\bm m\cdot\bm\theta:=\sum_{a=1}^d m_a\theta_a$. Sampling Eq.~\eqref{S:eq:multiF} along the physical lattice gives the wave-number magnitude
\begin{equation}
 k_{\bm m}=2\pi\normZ{\bm m\cdot\bm\beta}.
 \label{S:eq:multiq}
\end{equation}

Choose any fixed norm $\|\cdot\|$ on $\mathbb R^d$; all such norms are equivalent for the logarithmic exponent below. Define the dual resonance exponent
\begin{equation}
 \omega_*(\bm\beta):=\limsup_{\|\bm m\|\to\infty}
 \frac{-\log\normZ{\bm m\cdot\bm\beta}}{\log\|\bm m\|}.
 \label{S:eq:multiomega}
\end{equation}
For $d=1$, $\bm m$ reduces to an integer $m$, Eq.~\eqref{S:eq:multiomega} becomes Eq.~\eqref{S:eq:omega}, and hence
\begin{equation}
 \omega_*(\beta)=\mu(\beta)-1.
 \label{S:eq:oneFreqReduction}
\end{equation}
This is the precise sense in which $\omega_*$ generalizes the one-frequency irrationality exponent relevant to physical resonances.

Assume $\omega_*(\bm\beta)<\infty$ and let $r>d/2$. We require a global upper bound: for some $B>0$ and all sufficiently large $\|\bm m\|$,
\begin{equation}
 |c_{\bm m}|\le B\|\bm m\|^{-r}.
 \label{S:eq:multitailUpper}
\end{equation}
We also assume that there exist $A>0$ and a sequence $\bm q_j\in\Z^d$ with $\|\bm q_j\|\to\infty$ that realizes the limsup in Eq.~\eqref{S:eq:multiomega} and satisfies
\begin{equation}
 |c_{\bm q_j}|\ge A\|\bm q_j\|^{-r}.
 \label{S:eq:multitailLower}
\end{equation}
As in the one-frequency theorem, the lower bound is needed only on this exceptional subsequence. Define the multi-frequency integrated weight by
\begin{equation}
 Z(K):=\sum_{\substack{\bm m\in\Z^d\setminus\{\bm0\}\\
 \normZ{\bm m\cdot\bm\beta}<K}}
 |c_{\bm m}|^2.
 \label{S:eq:multiZ}
\end{equation}

\textbf{Theorem S2.} Under the assumptions above,
\begin{equation}
 \liminf_{K\to0^+}\frac{\log Z(K)}{\log K}
 =\frac{2r}{\omega_*(\bm\beta)}.
 \label{S:eq:multiTheorem}
\end{equation}
For a one-dimensional physical structure factor, defining $\alpha+1$ by the left-hand side gives
\begin{equation}
 \alpha+1=\frac{2r}{\omega_*(\bm\beta)}.
 \label{S:eq:multiAlpha}
\end{equation}

\textit{Proof.} Fix $\epsilon>0$. By the definition of Eq.~\eqref{S:eq:multiomega}, there is a radius $R_\epsilon$ beyond which
\begin{equation}
 \normZ{\bm n\cdot\bm\beta}\ge
 \|\bm n\|^{-(\omega_*+\epsilon)}
 \label{S:eq:multiLower}
\end{equation}
for every integer vector $\bm n\in\Z^d$ with $\|\bm n\|\ge R_\epsilon$. Rational independence of $1,\beta_1,\ldots,\beta_d$ implies that $\normZ{\bm n\cdot\bm\beta}>0$ for every nonzero $\bm n\in\Z^d$. Hence, the finitely many vectors with $0<\|\bm n\|<R_\epsilon$ have a strictly positive minimum mismatch. For sufficiently small $K$, every $K$-resonant index $\bm m$ must therefore satisfy $\|\bm m\|\ge R_\epsilon$, and Eq.~\eqref{S:eq:multiLower} gives
\begin{equation}
 \|\bm m\|\gtrsim
 L_K:=K^{-1/(\omega_*+\epsilon)}.
 \label{S:eq:multiFirst}
\end{equation}
If $\bm m$ and $\bm m'$ are distinct $K$-resonant Fourier indices, then
\begin{equation}
 \normZ{(\bm m-\bm m')\cdot\bm\beta}<2K.
\end{equation}
After excluding the same finite set of small difference vectors, Eq.~\eqref{S:eq:multiLower} gives
\begin{equation}
 \|\bm m-\bm m'\|\gtrsim L_K.
 \label{S:eq:multiSep}
\end{equation}
Thus the resonant set is $L_K$-separated and begins at radius of order $L_K$. Such a subset of $\Z^d$ contains at most $C_d j^{d-1}$ points in the annulus $jL_K\le\|\bm m\|<(j+1)L_K$, where $C_d$ is independent of $j$ and $K$. Using Eq.~\eqref{S:eq:multitailUpper}, we obtain
\begin{align}
 Z(K)
 &\le C\sum_{j\ge1}j^{d-1}(jL_K)^{-2r}\\
 &= C L_K^{-2r}\sum_{j\ge1}j^{d-1-2r}.
 \label{S:eq:multiUpper}
\end{align}
The series converges because $r>d/2$, yielding
\begin{equation}
 Z(K)\le C_\epsilon K^{2r/(\omega_*+\epsilon)}.
\end{equation}
This proves the lower bound $2r/\omega_*$ on the logarithmic exponent after $\epsilon\downarrow0$. Conversely, let $\{\bm q_j\}$ be the exponent-realizing sequence assumed in Eq.~\eqref{S:eq:multitailLower} and define
\begin{equation}
 K_j:=2\normZ{\bm q_j\cdot\bm\beta}.
\end{equation}
Equation~\eqref{S:eq:multitailLower} gives the single-mode bound
\begin{equation}
 Z(K_j)\ge A^2\|\bm q_j\|^{-2r},
\end{equation}
which provides the matching subsequential exponent $2r/\omega_*$. This proves Eq.~\eqref{S:eq:multiTheorem}.

As in Sec.~\ref{S:subsec:logNondegeneracy}, the fixed-amplitude lower bound on the exponent-realizing sequence can also be weakened without changing Eq.~\eqref{S:eq:multiTheorem}. Specifically, it is sufficient to require
\begin{equation}
\lim_{j\to\infty}
\frac{-\log |c_{\boldsymbol q_j}|}
{\log \|\boldsymbol q_j\|}
=r.
\label{S:eq:multiLogNondegeneracy}
\end{equation}
Indeed, defining
\[
b_j:=
-\frac{\log |c_{\boldsymbol q_j}|}{\log\|\boldsymbol q_j\|}
\to r,
\qquad
a_j:=
-\frac{\log\|\boldsymbol q_j\cdot\boldsymbol\beta\|_{\mathbb Z}}
{\log\|\boldsymbol q_j\|}
\to\omega_*(\boldsymbol\beta),
\]
the same single-mode argument gives
\[
\frac{\log Z(K_j)}{\log K_j}
\le
\frac{2b_j}
{a_j-\dfrac{\log2}{\log\|\boldsymbol q_j\|}}
\longrightarrow
\frac{2r}{\omega_*(\boldsymbol\beta)}.
\]
Thus, the logarithmic weakening of the nondegeneracy condition extends directly to the multi-frequency theorem.

\subsection{Multi-frequency stable-gap extension}
The stable-gap result requires additional analyticity of the Hamiltonian on the hull and a spectral gap that remains open under complex continuation. Assume that $H(\bm\theta)$ extends analytically to a complex polystrip $|\operatorname{Im}\theta_a|<\rho_0$ for every component $a=1,\ldots,d$, and that a spectral sector remains uniformly separated from the rest of the spectrum there. A Riesz contour $\Gamma$ with the same uniform meaning as in Eq.~\eqref{S:eq:Riesz} then gives an analytic spectral projector. If $d_{\bm m}$ denotes its Fourier coefficient at index $\bm m$, analyticity in the polystrip gives constants $C,\rho>0$ such that
\begin{equation}
 |d_{\bm m}|\le C e^{-\rho\|\bm m\|_1},
 \label{S:eq:multiExp}
\end{equation}
where $\|\bm m\|_1:=\sum_{a=1}^d|m_a|$ is the $\ell^1$ norm. From Eq.~\eqref{S:eq:multiLower}, a small-$K$ resonance must satisfy $\|\bm m\|\gtrsim K^{-1/(\omega_*+\epsilon)}$. The number of integer Fourier modes at radius $R$ grows only polynomially in $R$, while Eq.~\eqref{S:eq:multiExp} decays exponentially. Consequently, for every $p>0$, the deep-sector integrated weight obeys
\begin{equation}
 Z_{\rm deep}(K)=o(K^p)
\end{equation}
as $K\to0^+$. The shell hull profile is bounded and hence has square-summable Fourier coefficients. Now Cauchy--Schwarz inequality makes the shell--deep cross term superalgebraically small as well. Consequently, the stable-gap/Fermi-shell separation extends directly to a finite-dimensional torus hull.


\end{widetext}

\end{document}